\documentclass[aps,pra,amsmath,amssymb,twocolumn,showpacs,
nofootinbib 
]{revtex4-1}
\pdfoutput=1
\usepackage{epsfig}
\usepackage{subfigure}
\usepackage{amsmath, amssymb, amsthm, amsfonts, siunitx, youngtab, ytableau, color, esvect}
\usepackage{bm,mathtools,relsize}
\usepackage{braket}
\usepackage[english]{babel}
\usepackage[normalem]{ulem}
\usepackage{float}
\usepackage{graphicx,subfigure, natbib,multirow,float,dsfont}
\usepackage{hyperref}

\def\ket#1{|#1\rangle}
\def\bra#1{\langle#1|}

\def\Per{\mathrm{per}}

\newcommand{\imm}{{\rm imm}}
\newcommand{\Det}{{\rm det}}

\begin{document}
\title{Generalized multi--photon quantum interference}
\author{
	Max Tillmann$^1$,
	Si-Hui Tan$^2$,
	Sarah E. Stoeckl$^1$,
	Barry C. Sanders$^{3,4}$,
	Hubert de Guise$^5$,
	Ren{\'e} Heilmann$^6$,
	Stefan Nolte$^6$,
	Alexander Szameit$^6$,
	Philip Walther$^{1}$}
\affiliation{$^1$Faculty of Physics, University of Vienna, Boltzmanngasse 5, A-1090 Vienna, Austria}
\affiliation{$^2$Singapore University of Technology and Design, 20 Dover Drive, 138682 Singapore}
\affiliation{$^3$Institute for Quantum Science and Technology, University of Calgary, Alberta, Canada T2N 1N4}
\affiliation{$^4$Program in Quantum Information Science, Canadian Institute for Advanced Research, Toronto, Ontario M5G 1Z8, Canada}
\affiliation{$^5$Department of Physics, Lakehead University, Thunder Bay, Ontario, P7B 5E1, Canada}
\affiliation{$^6$ Institute of Applied Physics, Abbe Center of Photonics, Friedrich-Schiller Universit\"at Jena, Max-Wien-Platz 1, D-07743 Jena, Germany}

\begin{abstract}
Non--classical interference of photons lies at the heart of optical quantum information processing. This effect is exploited in universal quantum gates as well as in purpose--built quantum computers that solve the BosonSampling problem. Although non--classical interference is often associated with perfectly indistinguishable photons this only represents the degenerate case, hard to achieve under realistic experimental conditions. Here we exploit tunable distinguishability to reveal the full spectrum of multi--photon non--classical interference. This we investigate in theory and experiment by controlling the delay times of three photons injected into an integrated interferometric network. We derive the entire coincidence landscape and identify transition matrix immanants as ideally suited functions to describe the generalized case of input photons with arbitrary distinguishability. We introduce a compact description by utilizing a natural basis which decouples the input state  from the interferometric network, thereby providing a useful tool for even larger photon numbers.
\end{abstract}

\maketitle

\section{Introduction}
Interference is essential to many fields of physics. Remarkably this is not only tied to a wave description in the classical domain but holds also for the quantum regime when dealing with wavefunctions and probability amplitudes. Quantum interference was experimentally confirmed in impressive single--particle interferometry experiments carried out using electrons\cite{Hasselbach2010}, neutrons\cite{rauch2001neutron}, atoms\cite{Cronin2009} and molecules\cite{Hornberger2012,Eibenberger2013}. Quantum physics also allows two objects to interfere with each other. This two--particle interference is characterized by the second--order correlation function $G^{(2)}$ dating back to the pioneering work of Hanbury Brown and Twiss from the 1950s\cite{Brown1956}. Utilizing the bosonic nature of photons Hong, Ou and Mandel\cite{Hong1987} (HOM) performed a seminal $G^{(2)}$--measurement using single photons and a $50/50$ beam splitter. Initially intended as a precise measurement of the coherence time of the photons, their experiment is now at the heart of optical quantum metrology\cite{Giovannetti2011}, quantum computing\cite{OBrien2007, Aspuru-Guzik2012} and quantum communication\cite{Gisin2007}. Recently an intermediate model of quantum computing has refocused attention towards the findings of HOM. BosonSampling\cite{aaronson2011computational} utilizes even higher order correlations through the non--classical interference of a few dozen single--photons.

The recent development of quantum photonics technology\cite{OBrien2009} allows experiments using a growing number of photons and large, complex interferometric networks. Manipulating such large Hilbert--spaces requires well adapted tools in both theory and experiment. Although non--classical interference is often associated with perfectly indistinguishable photons this only represents the simplest case of photon states fully symmetric under permutation. Experimentally partial distinguishability is ubiquitous because the generation of indistinguishable multi--photon states currently remains a challenge. Moreover partial distinguishability is of fundamental interest highlighted by e.g. the nonmonotonicity of the quantum--to--classical transition\cite{Tichy2011,Ra2013}. In the following we present a novel description for the non--classical interference of multiple photons of arbitrary distinguishability propagating through arbitrary interferometers. We introduce a symmetry--adapted and therefore natural basis with basis states acting as the normal coordinates for the description of the non--classical interference of photons. In our perspective a different interferometer just depends on a different set of normal coordinates; the non--classical interference is determined solely by the properties of the photons. Distinguishability, as the central property, is tunable by treating temporal delay as an explicit parameter thereby allowing access to the whole spectrum of non--classical interference.

\section{Results}
\subsection{The quantum interference of two bosons}\hspace*{\fill} \\
In the case of two photons the Hong--Ou--Mandel dip has become a canonical implementation of an optical $G^{(2)}$--measurement. In this experiment two photons are injected into distinct input ports of a beam splitter, which is effectively an $m=2$ interferometer, where $m$ is the number of modes of the interferometer. One element of the output probability distribution corresponding to the case where the two photons exit the beam splitter in different output ports, is recorded via a coincidence measurement. In figure \textbf{1a} we show the coincidence probability $P_c$ that depends on the transformation matrix $B$, here defined by the splitting ratio of the beam splitter, and the distinguishability of the photons. In the prominent example of a balanced, i.e. $50/50$, beam splitter and perfectly indistinguishable photons the coincidence rate vanishes. The established technique to calibrate for the point of maximal non--classical interference relies on tuning the relative temporal delay $\Delta\tau$, i.e. the distinguishability between the two photons. This is described by an overlap integral which accounts for the key properties of the photons such as spectral shape, polarization, spatial mode in addition to the relative temporal delay.  The coincidence probability $P_c(\Delta\tau)$ in the general case corresponds to
\begin{align}
P_c(\Delta\tau) =& 
\int d\omega\int d\omega' |\bra{\psi_{in}}\hat{B}^{\dagger}\hat{a}^{\dagger}_1(\omega)\hat{a}^{\dagger}_2(\omega')\ket{0}|^2\nonumber\\
=&  \boldsymbol{v_2}^{\dagger} \big[\hat{R}^{(2)}(\Delta\tau)\big]\boldsymbol{v_2}\nonumber\\
=\begin{pmatrix}
	{\rm per}(B) \\
	{\rm det}(B)
	\end{pmatrix}^{\mathlarger{\dagger}}
	&\left[\frac{1}{2}
\begin{pmatrix}
	1 & 0 \\
	0 & 1
	\end{pmatrix}
+\frac{1}{2}\zeta\text{e}^{- \xi\Delta\tau^2}
\begin{pmatrix}
	1 & 0 \\
	0 & -1
	\end{pmatrix}
	\right]
\begin{pmatrix}
	{\rm per}(B) \\
	{\rm det}(B)
	\end{pmatrix}\label{eq:P112},
\end{align}
where $0 \leq \zeta \leq 1$ is derived from the mode--overlap integral, $\bra{\psi_{in}}=\bra{0}\hat{a}_1(\omega)\hat{a}_2(\omega')$ is the state impinging on the beamsplitter, and $\xi$ is a factor describing the shape of the interference feature (see supplementary information for further details).
\subsection{The natural basis for two photons}\hspace*{\fill} \\
In the case of two photons the non--classical interference is a second--order correlation effect and therefore dependent on the permutational symmetry of the two interfering particles. We consider a basis accounting for the permutational symmetries as a natural basis for quantum interference. Consequently we introduce a basis vector $\boldsymbol{v}$, whose components encapsulate the unitary network description $B$ in matrix functions having definite permutation properties; the first component is the permanent (per) and is fully symmetric under permutation; the second component is the determinant (det) and is fully antisymmetric under permutation. These are the only two possible symmetries when permuting two objects. By using the basis vector $\boldsymbol{v}$, we obtain an elegant and compact form of the rate matrix $\hat{R}^{(2)}(\Delta\tau)$; $\hat{R}^{(2)}(\Delta\tau)$ is a diagonal matrix and its entries depend only on properties of the input state. The ratio of its two non--zero entries, $\hat{R}^{(2)}_{11}$ and $\hat{R}^{(2)}_{22}$, are revealing the nature of the non--classical interference of two photons of arbitrary coherence. For indistinguishable photons and zero temporal delay $\Delta\tau$, $\hat{R}^{(2)}_{22}$ is also zero and the output probability is proportional to the permanent of $B$ only. The permutational symmetry of identical bosonic particles, e.g. photons, is reflected in transition amplitudes determined by a permutational symmetric function - the permanent. Temporal delays larger than the coherence time of the photons, $\Delta\tau\gg\tau_c$, result in complete loss of coherence. In this case, often characterized as classical behaviour of the photons, both matrix entries contribute equally, $\hat{R}^{(2)}_{11}=\hat{R}^{(2)}_{22}=0.5$. In this case the state is an equal mixture of symmetric and antisymmetric parts and so does not exhibit any of the indistinguishability features associated with quantum interference. The analysis above can be generalized to the quantum interference of two photons in larger interferometric networks. In this case the two input ports and the two ports in which the photons exit such a network define $2\times2$ scattering submatrices $B^*$. While the basis vector $\boldsymbol{v}$ now contains matrix functions of $B^*$ the rate matrix $\hat{R}^{(2)}(\Delta\tau)$ stays identical, independent of $B^*$. Figure \textbf{1b} highlights how this natural basis cleanly separates effects arising due to distinguishability in the input state from effects of the interferometric network. The advantage becomes increasingly evident for the non--classical interference of more than two photons.
\begin{figure}
\includegraphics[width=0.45\textwidth]{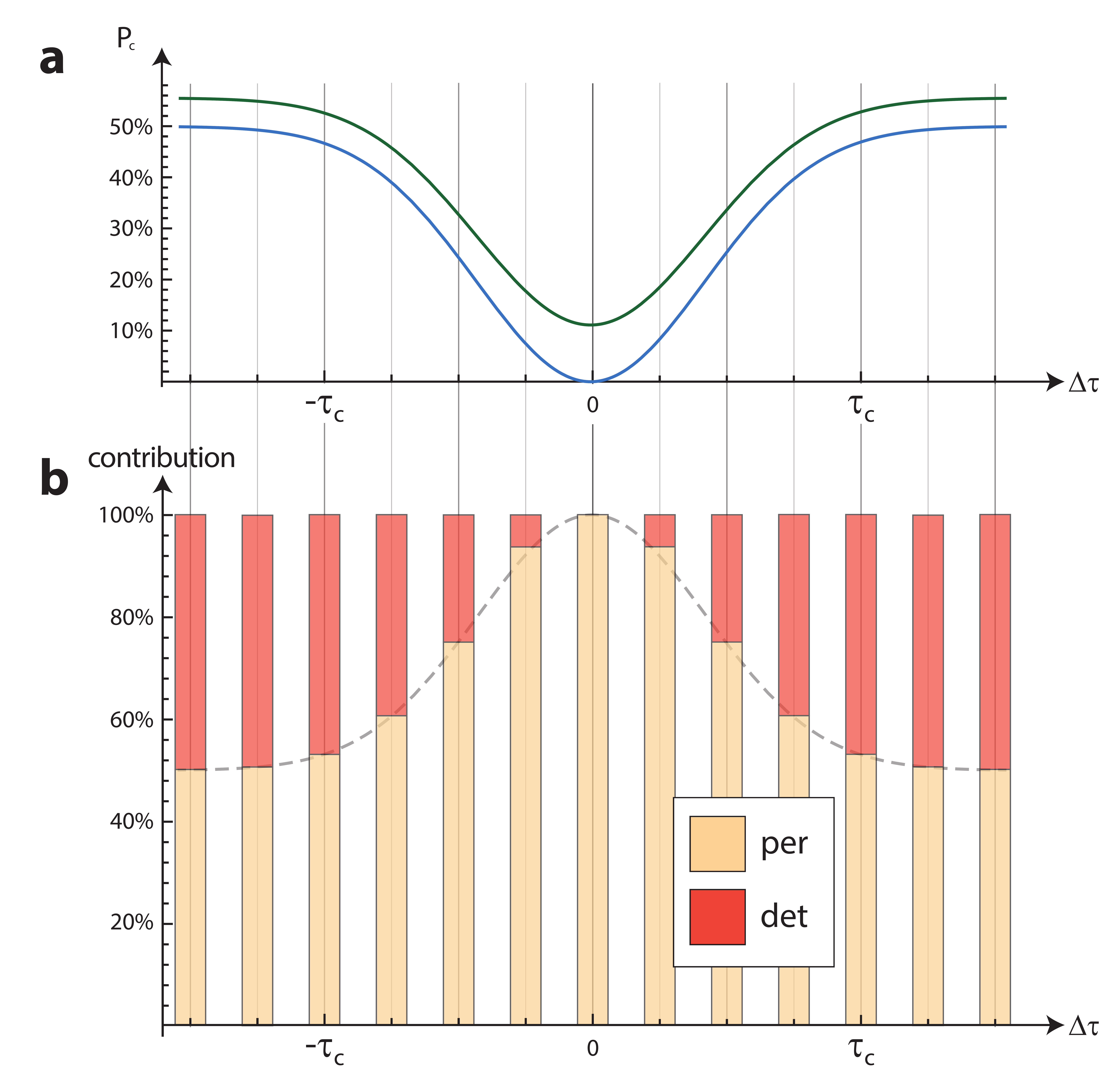}
\caption{\textbf{Two--photon non--classical interference.} Two photons of temporal coherence $\tau_c$ enter a beam splitter through different input ports. (\textbf{a}) The coincidence probability $P_c$ that they leave in two different output ports is plotted with respect to a relative temporal delay $\Delta\tau$. This delay is used to tune the distinguishability of the otherwise identical photons. The blue curve shows the output probability for a $50/50$ beam splitter, and the green one for a $67/33$ beam splitter. (\textbf{b}) depicts the contribution of the permanent (per) and determinant (det). It is the same for both beam splitters because this description is independent of the interferometer. In the case for zero delay ($\Delta\tau=0$) only the permanent contributes. By explicitly calculating the permanent, which is zero for a $50/50$ beam splitter, the vanishing output probability (\textbf{a}) for zero delay is obtained.}
\label{FIG:HOM}
\end{figure}

\subsection{The quantum interference of three bosons}\hspace*{\fill} \\
Consider a scenario where two photons are nearly indistinguishable and the third is delayed significantly. Adding a third photon leads to situations that can no longer be understood by the weighted sum of the permanent and determinant. In order to describe such a behaviour a more general matrix function, the immanant, is necessary\cite{Tan2013,Guise2014}. The immanant\cite{Littlewood1934} expands the concept of the permanent and determinant to mixed permutation symmetries and is defined as $\text{imm}(M)=\sum_{\sigma} \chi(\sigma)\prod_i M_{i\sigma(i)}$ for~$M_{ij}$ matrix elements of~$M$, with $\chi(\sigma)$ the character of permutation~$\sigma$. The permanent, for which every $\chi(\sigma)=1$, and the determinant, for which $\chi(\sigma)=\text{sgn}(\sigma)$, are special cases of the immanant (for an intuitive explanation of these matrix functions see figure \textbf{2}).

\begin{figure}
\includegraphics[width=0.45\textwidth]{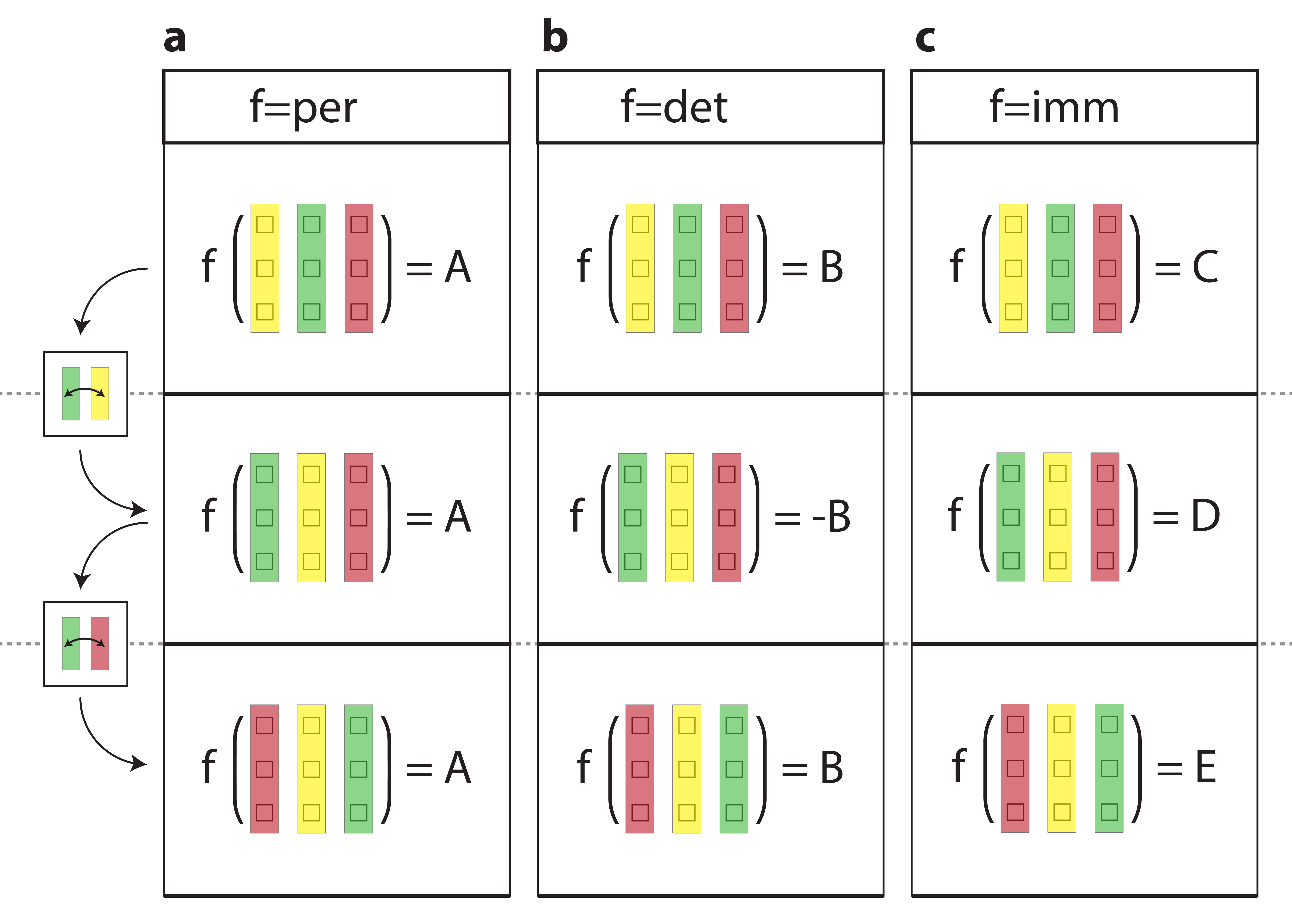}
\caption{\textbf{The permanent, the determinant and the immanant.} The permanent (per) and the determinant (det) are special cases of the immanant (imm), a matrix function. The three functions behave differently under odd permutations of the matrix. The permanent is symmetric under permutation of e.g. columns, depicted in (\textbf{a}). Permuting the yellow and green columns and calculating the permanent of this permuted and the original matrix yields the same result, A. Contrarily, the determinant of a permuted matrix will show a sign change compared to the determinant of the original matrix (\textbf{b}). Therefore the determinant is antisymmetric under odd permutations. Immanants (\textbf{c}) can describe both cases above, but their strength lies in covering mixed permutation symmetry. Calculating the immanant of a matrix and of another one with, e.g. the yellow and green columns flipped will give different results, C and D respectively. Additionally flipping the green and red column results in an even permutation of the original matrix. The immanant of this matrix gives yet another results, E. The overall number of immanants is bounded by the maximal number of unique permutation operations of the corresponding symmetric group $S_n$.}
\label{FIG:MatrixFunctions}
\end{figure}

In the smallest instance of a three--photon quantum interference $n=3$ photons are injected into a $m=3$--mode interferometric network and measured as three--fold coincidences at the three output ports. The optical transformation implemented by the interferometer can be any $3\times3$ linear optical transformation and the distinguishability of the three photons is arbitrarily tunable by setting the relative temporal delays; $\Delta\tau_1$ between the first and second photon and $\Delta\tau_2$ between the second and third photon.\\
The coincidence probability $P_{111}(\Delta\tau_1,\Delta\tau_2)$ is given in equation~(\ref{eq:P111}), where $\hat{a}^{\dagger}_1(\omega)$, $\hat{a}^{\dagger}_2(\omega')$ and $\hat{a}^{\dagger}_3(\omega'')$ are the creation operators in  modes $1,2,3$ of $T$ for photons with different spectral shape functions dependent on the frequency variables $\omega,\omega',\omega''$. Here $\bra{\psi_{in}}=\bra{0}\hat{a}_1(\omega)\hat{a}_2(\omega')\hat{a}_2(\omega'')$ is the three--photon state impinging on the interferometer.
An expression of equation~(\ref{eq:P111}), expanded in terms of immanants, determinants and permanents, results in a linear superposition of 60 terms. However, utilizing a symmetry--adapted basis allows for the compact representationso given in equation~(\ref{eq:P1112}) \&~(\ref{eq:P1113}) (see methods and supplementary information for further details). Here four immanants, the permanent and the determinant of $T$ constitute the components of a six--dimensional basis vector $\boldsymbol{v_3}$. The basis transformation $\hat{P}$ and $\hat{S}$, a matrix mapping matrix elements to matrix functions, transform between the basis vector of equation~(\ref{eq:P1112}), $\boldsymbol{a}=\hat{P}\hat{S}\boldsymbol{v_3}$, and the basis vector of equation~(\ref{eq:P1113}), $\boldsymbol{v_3}$. Analogous to equation~(\ref{eq:P112}) the $\zeta$ terms are derived from the mode--overlap integral and the $\xi$ terms are factors describing the shape of the interference feature. In this notation the overlap terms weight a sum of six matrices: the identity matrix and five permutation matrices $\rho_{12}$,$\rho_{13}$,$\rho_{23}$,$\rho_{123}$ and $\rho_{132}$, the subscripts of which label the permutation operation.
\subsection{The natural basis for three photons}\hspace*{\fill} \\
Whereas in equation~(\ref{eq:P111}) the basis states exhibit no particular permutation symmetry, states of the natural basis introduced to yield the fully block--diagonal form of equation~(\ref{eq:P1112}) have specific permutation properties: states of one symmetry type transform to states of the same type under permutation, i.e. they are decoupled under permutation. States in the natural basis thus play the role of normal coordinates for the non--classical interference of photons. Where equation~(\ref{eq:P1112}) highlights the six different permutational possibilities for three photons summing the matrices inside the square--brackets yields the $6\times6$ rate matrix $\hat{R}^{(3)}(\Delta\tau_1,\Delta\tau_2)$ of equation~(\ref{eq:P1113}). $\hat{R}^{(3)}(\Delta\tau_1,\Delta\tau_2)$ contains all the information regarding the input state, i.e. mode--mismatch and temporal delay to specify the non--classical interference of three photons independent of the scattering transformation $T$. Two entries of the block--diagonal rate matrix are sufficient for an interpretation. $F_{per}=\hat{R}^{(3)}_{11}(\Delta\tau_1,\Delta\tau_2)$ quantifies the fraction of the output probability distribution proportional to the permanent; the corresponding basis state is fully symmetric under permutation. $F_{det}=\hat{R}^{(3)}_{66}(\Delta\tau_1,\Delta\tau_2)$ quantifies the fraction of the output probability distribution proportional to the determinant; the corresponding basis state is fully antisymmetric under permutation. The contribution proportional to immanants can also be explicitly calculated. When only interested in their overall contribution it is given as $F_{imm}=1-F_{per}-F_{det}$. In the case for perfectly overlapping photons $F_{per}=1$ and therefore only the permanent of the scattering matrix contributes to the output probability distribution. Classical behaviour of the photons can be identified for $F_{per}=F_{det}=\frac{1}{6}$. As in the two photon case the input state and the interferometer decouple in the natural basis. As a consequence the treatment of the quantum interference of three photons in larger interferometric networks consisting of many modes becomes very efficient. For such a problem it is sufficient to calculate the rate matrix $\hat{R}^{(3)}$ only once. The scattering matrix $T$, necessary to calculate the basis vector $\boldsymbol{v_3}$ for a specific element of a output probability distribution is just a $3\times3$ submatrix of the larger scattering matrix. It is specified by the input ports of the photons and the ports in which they exit the interferometer. To obtain multiple elements of a probability distribution it is sufficient to determine their respective basis vector $\boldsymbol{v_3}$.
\begin{widetext}
\begin{align}
 P_{111}(\Delta\tau_1,\Delta\tau_2)= 
&\int d\omega\int d\omega' \int d\omega''|\bra{\psi_{in}}\hat{T}^{\dagger}\hat{a}^{\dagger}_1(\omega)\hat{a}^{\dagger}_2(\omega')\hat{a}^{\dagger}_3(\omega'')\ket{0}|^2 \label{eq:P111}\\
 = &(\hat{P}\hat{S}\boldsymbol{v_3})^{\dagger} \big[{1\!\!1} + \rho_{12}\zeta_{12}\text{e}^{-\xi_{12}\Delta\tau_1^2}+ \rho_{23}\zeta_{23}\text{e}^{-\xi_{23}\Delta\tau_2^2}
 + \rho_{13}\zeta_{13}\text{e}^{-\xi_{13}(\Delta\tau_1-\Delta\tau_2)^2}\nonumber\\ & + \zeta_{123}\left(\rho_{132}\text{e}^{\xi_{123}^*(\Delta\tau_1,\Delta\tau_2)}+\rho_{123}\text{e}^{\xi_{123}(\Delta\tau_1,\Delta\tau_2)}\right)\big](\hat{P}\hat{S}\boldsymbol{v_3}),\label{eq:P1112}\\
= &\boldsymbol{v_3}^{\dagger}\big[\hat{R}^{(3)}(\Delta\tau_1,\Delta\tau_2)\big]\boldsymbol{v_3}\label{eq:P1113},
\end{align}
\end{widetext}

\begin{figure*}
\includegraphics[width=0.925\textwidth]{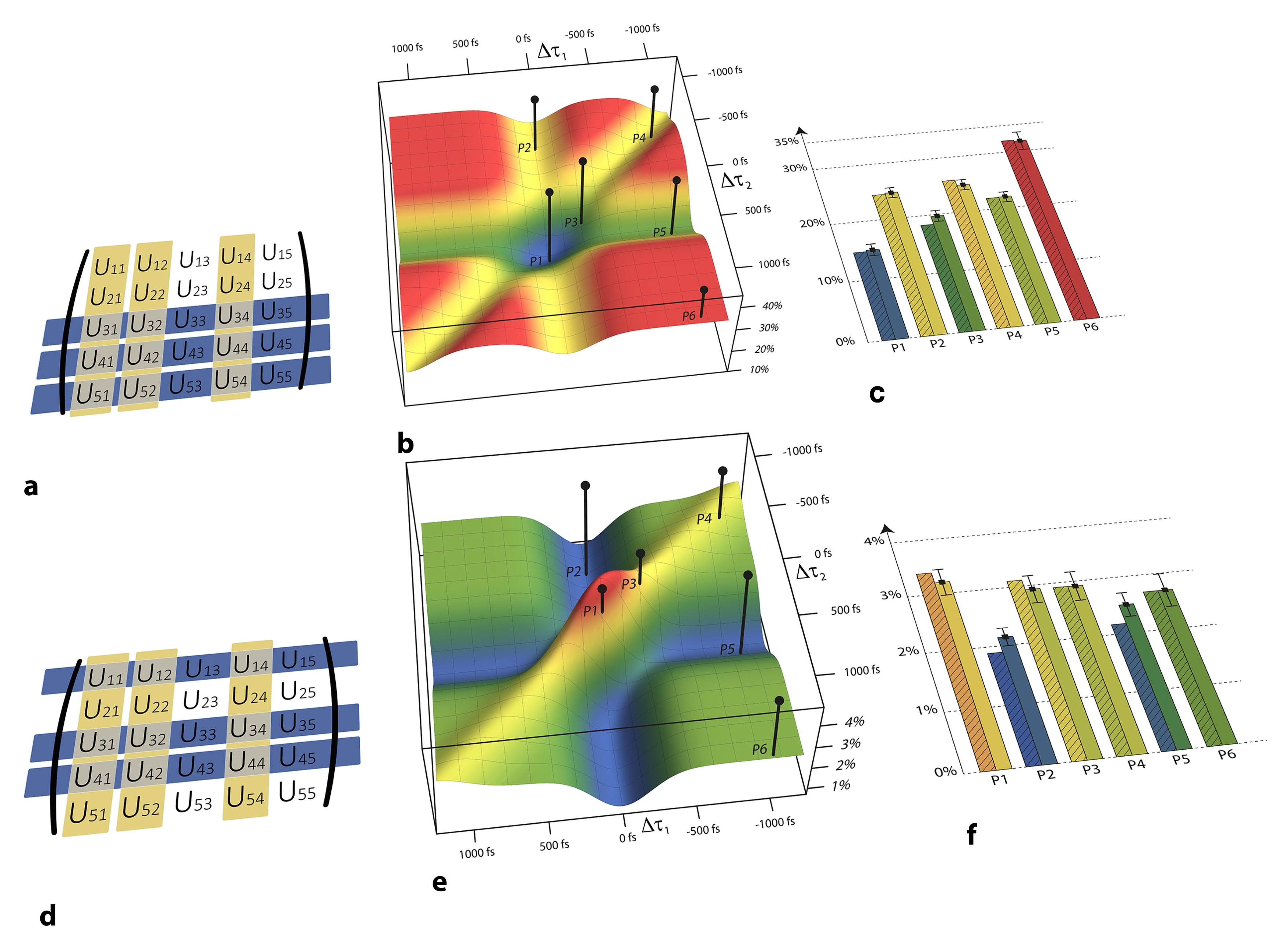}
\caption{\textbf{Three--photon coincidence landscapes} Three photons enter the interferometric network, one each in mode 1, 2 and 4 (highlighted in yellow), and exit the network in modes \textbf{a)} 3, 4 and 5 and \textbf{d)} 1, 3 and 4 respectively (highlighted in blue). The intersections of the inputs' columns and the outputs' rows uniquely select matrix entries that constitute $3\times3$ submatrices. Tuning the temporal delay of the three photons with respect to each other ($\Delta\tau_1$ and $\Delta\tau_2$) gives rise to coincidence landscapes (\textbf{b)} and \textbf{e)}). Their temporal distinguishability determines the degree of non--classical interference and therefore probability to detect such an event. Six characteristic points (P1 ... P6) of each landscape are experimentally sampled. Theoretical prediction (left bars, shaded) and experimentally obtained output probabilities (right bars) for the six points and both output combinations are shown in \textbf{c)} and \textbf{f)}. The reduced $\chi^2$ is 1.38 and 1.10 respectively and the experimental errors are calculated as standard deviations.}
\label{FIG:GBS}
\end{figure*}

\subsection{The coincidence landscape}\hspace*{\fill} \\
In the experiment four--photon events generated by higher--order emission from a spontaneous parametric down--converter are distributed to four different spatial modes. Relying on a detection event in the trigger mode and post--selection, the three--photon input state, one photon in each input mode coupled to the interferometer, is heralded. We ensure that all photons are indistinguishable in a polarization basis. The spectral properties of these photons are independently measured using a single--photon spectrometer. Their relative temporal delay $\Delta\tau_1$ and $\Delta\tau_2$ can be set using motorized delay lines. The transformation of the fs--written integrated interferometer, a $5\times5$ unitary matrix, is recovered using the reconstruction method specified in the supplementary material. Injecting the photons in three input ports of the interferometer and detecting them in three separate output ports uniquely selects a $3\times3$ scattering submatrix $T$ (see figure \textbf{3a} and \textbf{3d}). For each $3\times3$ submatrix, using a precisely tunable delay allows us to reveal the full spectrum and thereby nature of the non--classical interference. We visualize this as a three--dimensional coincidence landscape as shown in figure \textbf{3b} and \textbf{3e}. The relief of such a landscape features distinct "landforms" are in correspondence with distinguishability features of the photons. In the center region, $\Delta\tau_1=\Delta\tau_2=0\pm\tau_c$, a peak or dip arises due to constructive or destructive interference of all three photons. Note that the absolute zero position $\Delta\tau_1=\Delta\tau_2=0$ corresponds to a permanent only in the absence of any spectral distinguishability. Along the three axis $\Delta\tau_1=0$, $\Delta\tau_2=0$ and $\Delta\tau_1=\Delta\tau_2>|\pm\tau_c|$ valleys or ridges form due to the non--classical interference of two indistinguishable photons with the third one being distinguishable. Each valley or ridge depicts a case where one of the three photons is distinguishable compared to the other two photons. Along those ridges and valleys the output probability is largely proportional to immanants of the scattering matrix. "Classical" behaviour, i.e. complete distinguishability, of the three photons is associated to plateaux for $\Delta\tau_1=-\Delta\tau_2>|\pm\tau_c|$. These are the only areas where determinants of the scattering matrix contribute, accounting for the anti--symmetrical part of the input state. Coincidences for six points of pairwise different temporal delays, $P1$ to $P6$, for two different scattering submatrices (see figure \textbf{3c} and \textbf{3f}) are measured. These six points were selected because they highlight the connection between landscape features, permutation symmetries, and partial distinguishability. Furthermore they provide a sufficient set of experimental data for fitting the coincidence landscapes. A reduced $\chi^2$ of $1.38$ and $1.10$ for the two landcapes quantifies the overlap between our theory and the experiment. The deviations are most likely due to higher--order emissions and frequency correlations of the input state.\\
The landscape interpretation can be extended to the interference of larger numbers of photons $n$, which generate n-dimensional "hyperlandscapes". These are spanned by $n$-$1$ axes of pairwise temporal delays with the last axis representing the actual coincidence rate. The "landforms" reach from complex $n$--dimensional features corresponding to the partial indistinguishability of all $n$ photons to the simple one-dimensional plateaux associated with completely distinguishable photons.

\begin{figure*}
\includegraphics[width=0.90\textwidth]{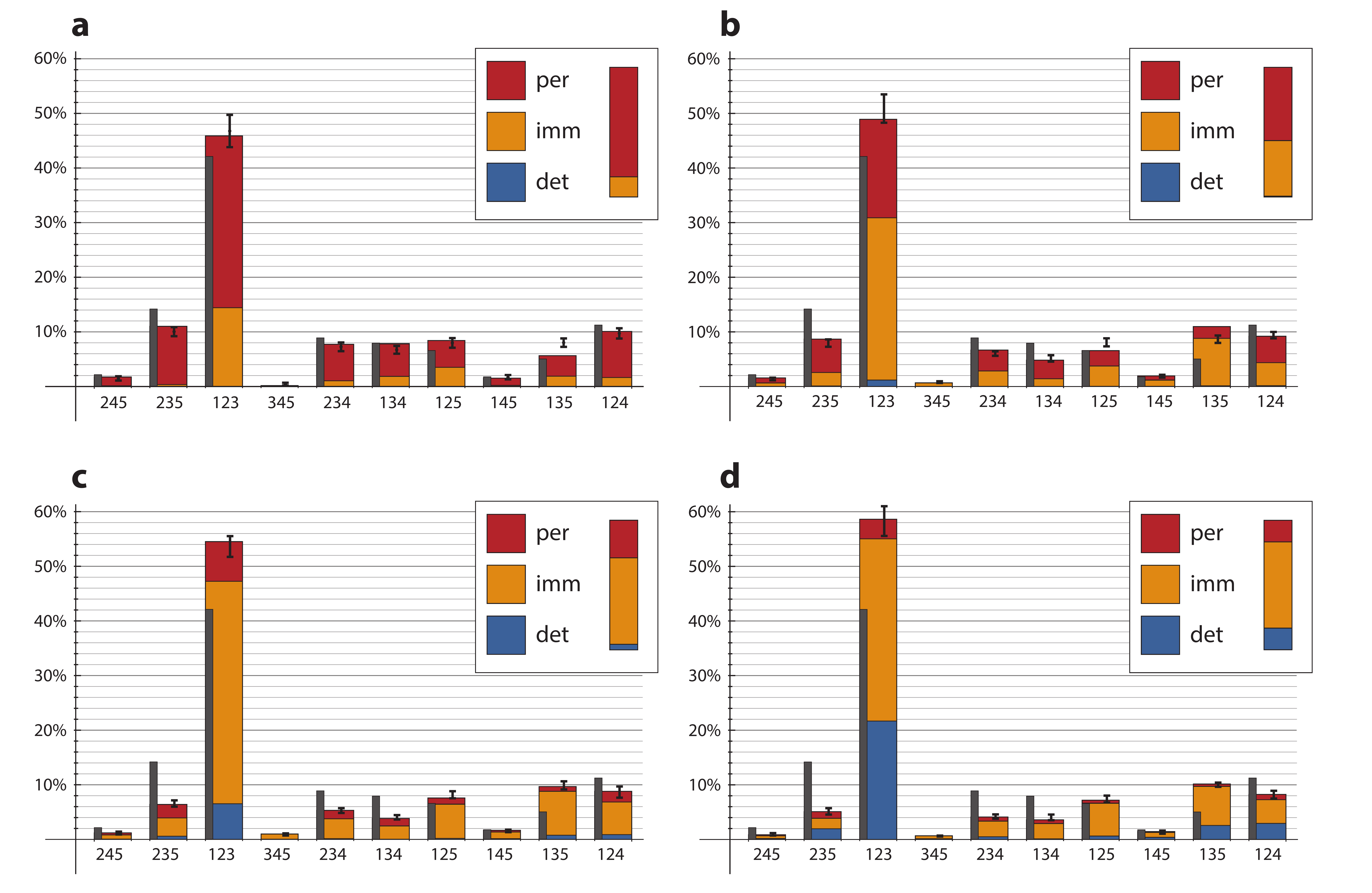}
\caption{\textbf{Experimental BosonSampling output distribution for in--, semi-- and distinguishable photons} Various temporal delays for three photons lead to different contributions of permanents, immanants, and determinants. The normalized output probability distributions for the three photons is measured as coincidences from different spatial modes, resulting in ten elements. For all temporal delays the three photons exhibit slight spectral mismatch. \textbf{a} depicts the case for a small temporal offset $\Delta\tau_1<<\tau_c>>\Delta\tau_2$ (P1 of figure 3), whereas for \textbf{b} and \textbf{c} this delay is increased along a diagonal axis $\Delta\tau_1 \approx \Delta\tau_2$ (P3 and P4 of figure \textbf{3}). The extreme case of complete distinguishability and therefore classical behavior is shown in \textbf{d} (P6 of figure \textbf{3}). As a reference the grey bars illustrate the case for perfect indistinguishability and therefore only contribution from the permanent of the scattering submatrix. The interferometer independent contribution $F_{per}$, $F_{det}$ and $F_{imm}$ is shown in the figure legend. The error bars of the experimental data are standard deviations over 19 independent runs.}
\label{FIG:BosonSampling}
\end{figure*}

\subsection{From permanents to immanants}\hspace*{\fill}\\
Quantum computing leverages quantum resources to efficiently perform certain classically hard computations\cite{Nielsen2010}. Whereas many quantum algorithms solve a certain decision problem, BosonSampling introduces a new paradigm: it seeks efficient sampling of a distribution of matrix transformations, which is a task hard to implement efficiently on classical computers. Optical realizations of both approaches, universal quantum computing and BosonSampling, rely intrinsically on the non--classical interference of more than two photons. BosonSampling is singular amongst current proposals because of its low requirements of space and time resources, brings within current technological reach the realistic possibility of demonstrating the superiority of quantum computing. This promise has led to several BosonSampling experiments\cite{Broome2013,Spring2013,Tillmann2013,Crespi2013} and follow--up work\cite{Gogolin2013,Aaronson2013,Spagnolo2014,Carolan2014}.\\ In order to scale BosonSampling to larger instances two main issues need to be addressed. The first issue is the technology\cite{Lita2008,Zhou2014,Marsili2013} needed to increase the size of the instances implemented. The second issue is handling of possible errors\cite{Leverrier2013,Rohde2012,Rohde2012a}. BosonSampling is a purely passive optical scheme and therefore lacks error correction capabilities\cite{Rohde2014a}. The success of computation depends crucially on the quality of the experimental apparatus. Only in the ideal case where the interfering photons are indistinguishable in all degrees of freedom is the resulting output probability distribution proportional to the permanent only. Our analysis exposes that this condition is rather fragile and therefore distinguishability must be regarded as the dominant source of error. \\
Remarkably, large classes of immanants are known to be in the same complexity class as permanents\cite{burgisser2000computational,brylinski2003complexity}. Thus it is an intuitive conjecture that output probability distributions depending largely on immanants rather than just the permanant are also computationally hard. Whether this holds for sampling from these distributions is an active field of research. \\
Optical implementations of BosonSampling instances utilize state--of--the--art large--scale random scattering networks and generate huge output probability distributions consisting of many elements. These prerequisites make it a benchmark for multi--photon non--classical interference. Consequently a description of generalized non--classical interference needs to be assessed under these conditions. Our approach decouples the interferometer from the non--classical interference hence the treatment and conclusions become analogue for e.g. central building blocks of linear optical quantum computing like ancillae assisted CNOT--gates. These typically feature more symmetric and simpler networks however, rendering the non--classical interference far less rich.

\ytableausetup{mathmode, boxsize=3pt}
\begin{figure*}
\includegraphics[width=0.95\textwidth]{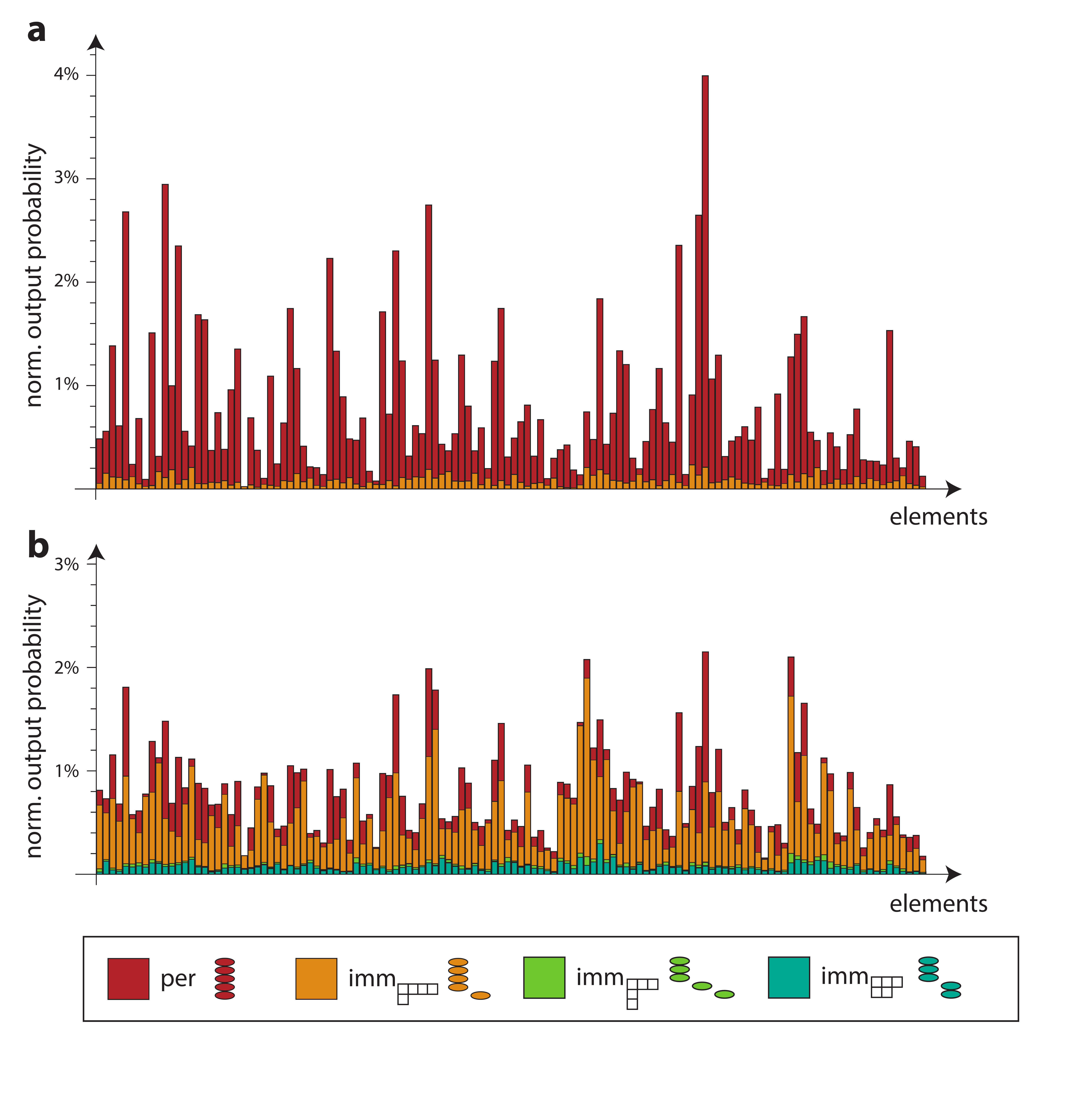}
\caption{\textbf{Five--photon BosonSampling including distinguishability:}Simulation of a BosonSampling instance of five photons propagating through an interferometric network of nine modes. For different degrees of photon distinguishability (\textbf{a}, \textbf{b}) contributions from permanents (per) and immanants (imm$_{\ydiagram{4,1}}$, imm$_{\ydiagram{3,1,1}}$, imm$_{\ydiagram{3,2}}$) arise. The immanant contributions cover physical scenarios with different symmetries under exchange of five photons. The labels of the immanants describe the number of photons that are distinguishable. imm$_{\ydiagram{4,1}}$, for instance, is the contribution from the case when four photons are indistinguishable from one another but distinguishable from the fifth photon. The output probability distribution of five photons exiting the interferometer in five different modes is normalized and contains 126 elements. \textbf{a} depicts the close to the ideal case where realistic errors such as slight spectral mismatch and temporal delay ($\Delta\tau_i\leq\frac{1}{20}\tau_c$) of the photons lead to a small degree of partial distinguishability. \textbf{b} shows a case where the interfering photons exhibit increased partial distinguishability ($\Delta\tau_i\leq\frac{1}{5}\tau_c$).}
\label{FIG:BosonSampling_sim}
\end{figure*}

\subsection{Investigation of a BosonSampling computer}\hspace*{\fill} \\
We investigate generalized non--classical interference of three photons in a five--moded interferometric network in theory and experiment. This serves a dual purpose: On one hand it emphasizes how distinguishability influences a three--photon BosonSampling instance. On the other hand the full permutational spectrum of a generalized non--classical interference is shown for complex networks exhibiting a random structure. The photons exhibit some spectral mismatch and are additionally rendered fully or partially distinguishable by controlling temporal delays. Figure \textbf{4a} illustrates the result for partial distinguishability, whereas in figure \textbf{4b} and \textbf{4c} this distinguishability is increased by varying the temporal delay along a diagonal axis $\Delta\tau_1\approx\Delta\tau_2$. The extreme case of complete distinguishability and thus classical behaviour is shown in figure \textbf{4d}. As reference we include in all figures the ideal case of zero delay and perfect indistinguishability as grey bars. The interferometer independent contribution $F_{per}$, $F_{det}$ and $F_{imm}$ is contained as an inset in the legend of each figure.

The elements of each output probability distribution are recovered by calculating the corresponding matrix functions. Note that for each element the absolute value of these matrix functions, e.g. $|\text{per}(T)|^2$ or $|\text{det}(T)|^2$, can vary largely depending on the scattering submatrix $T$. This is pronounced for the output event $123$ where $|\text{per}(T_{123})|^2\approx\frac{1}{5}|\text{det}(T_{123})|^2$. In general the fraction of the output probability distribution proportional to the permanent drops rapidly with increasing distinguishability. Instead contributions from immanants become dominant and reflect cases where only two of the three photons interfere non--classically. For large delays along the antidiagonal axis $\Delta\tau_1\approx-\Delta\tau_2$ the three photons' wavefunctions do not overlap anymore and the determinant contributes with $F_{det}=\frac{1}{6}$ (see figure \textbf{4d}). For comparably large delays along the diagonal axis $\Delta\tau_1\approx\Delta\tau_2$ two photons stay nearly indistinguishable and the contribution from the determinant is suppressed to $F_{det}\approx0$ (see figure \textbf{4c}). The classical case (figure \textbf{4d}) can be always identified with equal contribution from the permanent and determinant $F_{per}=F_{det}=\frac{1}{n!}$, which is for $n=3$ photons $F_{per}=F_{det}=\frac{1}{6}$.\\
Our theory emphasizes the permutation symmetries of $n$ photons using the representation theory of the symmetric group $S_n$. The theory is thus independent of the number of modes $m$ in the interferometer, a feature that is extremely convenient for large scale networks where $m\gg n$, even though the representations increase with with $n!$. In figure \textbf{5} we show the applicability of our method for larger $n$ and $m$ with a calculation of a generalized non--classical interference of five photons injected in a network of nine modes. The full spectrum of such a non--classical interference, constituted by permanents, determinants and immanants of the respective scattering submatrices is revealed by tuning the photons' distinguishability. Different physical scenarios of partial distinguishability, e.g the case where four photons are indistinguishable from one another but distinguishable from the fifth photon, are covered by the corresponding partitions of the immanants. Figure \textbf{5b} highlights that already partial distinguishability significantly alters the output probability distribution to be primarily proportional to immanants.

\section*{Discussion}
We present a novel analysis of multi–photon quantum interference revealing the full permutational spectrum of input states with arbitrary distinguishability. A comprehensive physical interpretation is achieved/given by establishing a correspondence between matrix immanants and these mixed symmetry input states. We introduce a rate-matrix containing all the information on the non-classical interference and basis vectors containing the information on the interferometric network. Output probabilities are recovered as an inner product of these vectors with the rate-matrix serving as a metric. This rate-matrix is block-diagonalized and each block corresponds to a different physical scenario of non-classical interference.
This indicates that this block diagonalization and consequent interpretation are not only fundamental but also universal features of multi-photon interferometry. We experimentally confirm our theory by recovering
the full coincidence landscape of three arbitrarily distinguishable photons and give an analytical example for five photons. Our approach thus provides a deeper understanding of the rich spectrum of multi-photon non-classical interference. Additionally our method can be used to characterize a broad range of optical interferometers used for example in quantum information processing. While passive schemes like BosonSampling benefit most from this approach it applies analogously to crucial building blocks of linear optical quantum computing relying on the non--classical interference of more than two photons\cite{Knill2001,Pittman2001}.
\\

\section{Methods}
\textbf{Three--photon coincidence probability}
Vector $\hat{P}\hat{S}\boldsymbol{v}$ in equation~(\ref{eq:P1112}) is defined as:
\begin{widetext}
\begin{align}
\hat{P}\hat{S}\boldsymbol{v}=\left(\begin{array}{c}
\text{per}(T) \\
\text{det}(T) \\
\frac{1}{2\sqrt{3}}\text{imm}(T)+\frac{1}{2\sqrt{3}}\text{imm}(T_{312})\\
\frac{1}{6}\text{imm}(T)-\frac{1}{3}\text{imm}(T_{132})-\frac{1}{6}\text{imm}(T_{213})+\frac{1}{3}\text{imm}(T_{312})\\
\frac{1}{6}\text{imm}(T)+\frac{1}{3}\text{imm}(T_{132})+\frac{1}{6}\text{imm}(T_{213})+\frac{1}{3}\text{imm}(T_{312})\\
-\frac{1}{2\sqrt{3}}\text{imm}(T)+\frac{1}{2\sqrt{3}}\text{imm}(T_{213})\ ,
\end{array}\right )
\end{align}
\end{widetext}
where $T_{ijk}$ is the matrix $T$ in which rows 1,2 and 3 have been rearranged in order $i$, $j$, $k$.
\\

\begin{figure*}[ht!]
\includegraphics[width=0.95\textwidth]{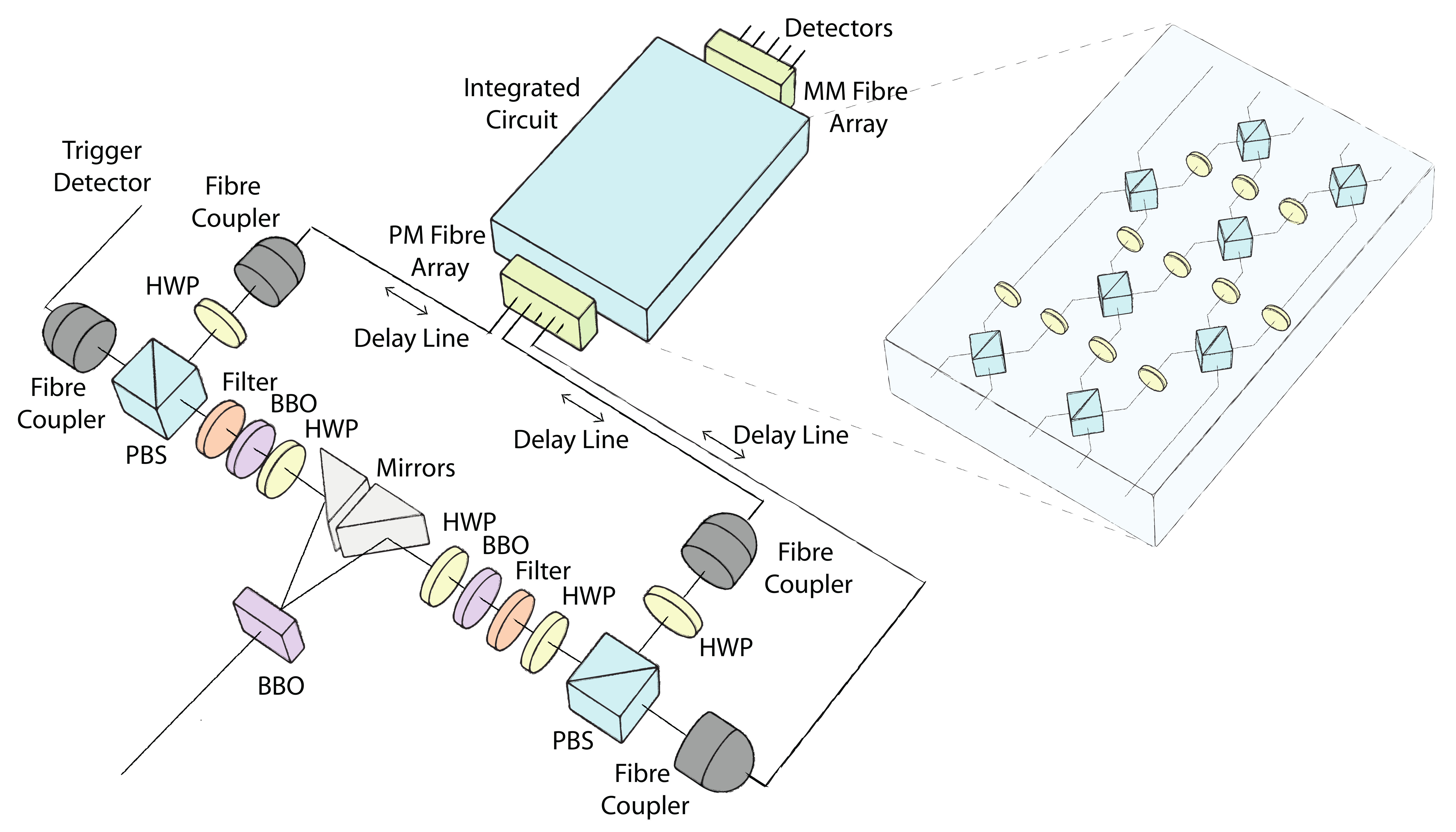}
\caption{\textbf{Experimental setup.} Four photons are generated via spontaneous parametric down--conversion (SPDC) and distributed to four spatial modes with two PBSs. A four--fold coincidence event consisting of three photons exiting the network and a trigger event postselects the desired input state. The delay lines allow to tune the distinguishability and therefore quantum--interference of the three photons propagating through the waveguide. The integrated circuit is shown in a Mach--Zehnder--decomposition and consists of eight beam splitters and eleven phase shifters.}
\label{Fig:Setup}
\end{figure*}
\newpage
\textbf{Labels of immanants by Young diagrams}
The different immanants in the caption of Fig. $5$ are indexed with the corresponding young diagrams. The Young diagrams are a collection of boxes here used to distinguish different physical scenarios of multi--photon non--classical interference. Young diagrams are a pictorial representation of different partitions of $S_n$\cite{Guise2014}.\\[\baselineskip]

\textbf{State generation}
A Ti--Sapphire oscillator emitting \SI{150}{\femto\second} pulses at \SI{789}{\nano\meter} and a repetition rate of \SI{80}{\mega\hertz} is frequency doubled in a $\text{LiB}_3\text{O}_5$ (LBO) crystal (see Fig.~\ref{Fig:Setup} for a schematic of the experimental setup). The output power of this second harmonic generation can be controlled by a power regulation stage consisting of a half--wave plate (HWP) and a polarizing beam splitter (PBS) placed before the LBO-crystal. The resulting emission at \SI{394.5}{\nano\meter} is focused into a \SI{2}{\milli\meter} thick $\beta\text{--BaB}_2\text{O}_4$ (BBO) crystal cut for degenerate non--collinear type--II down--conversion\cite{Kwiat1995}. A compensation scheme consisting of HWPs and \SI{1}{\milli\meter} thick BBO--crystals is applied for countering temporal and spatial walk--off. The two spatial outputs of the down--converter pass through narrowband interference filters ($\lambda_{\text{FWHM}}=$\,\SI{3}{\nano\meter}) to achieve a coherence time greater than the birefringent walk--off due to group velocity mismatch in the crystal ($|v_{g_e}-v_{g_o}|$ $\times$ half crystal thickness). Additionally this renders the photons close to spectral indistinguishability. The down--conversion--source is aligned to emit the maximally entangled Bell--state $\ket{\phi^+}=\frac{1}{\sqrt{2}}\left(\ket{HH}+\ket{VV}\right)$ when pumped at \SI{205}{\milli\watt} cw--equivalent pump power. The state is coupled into single mode fibers (Nufern 780--HP) equipped with pedal--based polarisation controllers to counter any stress--induced rotation of the polarisation inside the fiber. Each of these spatial modes is then coupled to one input of a PBS while its other input is occupied with a vacuum--state. The outputs pass HWPs and are subsequently coupled to four polarisation maintaining fibers (Nufern PM780--HP). Temporal overlap is controlled by two motorized delay lines that exhibit a bidirectional repeatability of $\pm\,$\,\SI{1}{\micro\meter}. Temporal alignment precision is limited by other factors in the setup to approximately $\pm\,$\,\SI{5}{\micro\meter} and is therefore within a precision of 2.5\,\% of the coherence length of the photons. The polarisation maintaining fibers are mated to a single mode fiber v--groove--array (Nufern PM780--HP) with a pitch of \SI{127}{\micro\meter} and butt--coupled to the integrated circuit. The coupling is controlled by a manual six--axis flexure stage and stable within 5\,\% of the total single--photon counts over 12 hours. The output fiber array consists of a multimode v--groove--array (GIF--625) and the photons are detected by single--photon avalanche photodiodes which are recorded with a home--built Field Programmable Gate Array logic. The coincidence time window was set to \SI{3}{\nano\second}.
\\In order to measure the six points of the coincidence landscapes a three--photon input state was injected into the integrated network (see supplementary information for further details). Therefore the BBO was pumped with cw-equivalent power of \SI{700}{\milli\watt} and the ratio of the six--photon emission over the desired four--photon emission was measured to be below 5\,\%. 
\\[\baselineskip]
\textbf{Integrated network fabrication.}
The integrated photonic networks were fabricated using a femto\-second\ direct\---write writing technology\cite{Itoh2006,Marshall2009}. Laser pulses were focused \SI{370}{\micro\meter} below the surface of a high--purity fused silica wafer by an NA\,=\,0.6 objective. The \SI{200}{\nano\joule} pulses exhibit a pulse duration of \SI{150}{\femto\second} at \SI{100}{\kilo\hertz} repetition rate and a central wavelength of \SI{800}{\nano\meter}. In order to write the individual waveguides the wafer was translated with a speed of \SI[per-mode=symbol]{6}{\centi\meter\per\second}. The waveguide modes exhibit a mode field diameter of \SI{21.4}{\micro\meter} $\times$ \SI{17.2}{\micro\meter} for a wavelength of \SI{789}{\nano\meter} and a propagation loss of \SI[per-mode=symbol]{0.3}{\decibel\per\centi\meter}.
This results in a coupling loss of \SI{-3.5}{\decibel} with the type of input fibers used in this experiment.
Coupling to the output array results in negligible loss due to the use of multimode fibers. 
\\[\baselineskip]
\vspace*{10cm}
\vspace*{4\baselineskip}

\begin{acknowledgments}
The authors thank I. Dhand and J. Cotter for helpful discussions, M. Tomandl for assistance with the illustrations and J. Nielsen and J. Kulp for assistance with coding and computing the five--photon non-classical interference. M.T., S.E.S. and P.W. acknowledge support from the European Commission with the project  EQuaM -Emulators of Quantum Frustrated Magnetism (No 323714), GRASP - Graphene-Based Single--Photon Nonlinear Optical Devices (No 613024), PICQUE - Photonic Integrated Compound Quantum Encoding (No 608062), QuILMI - Quantum Integrated Light Matter Interface (No 295293) and the ERA-Net CHIST-ERA project QUASAR - Quantum States: Analysis and Realizations, the German Ministry of Education and Research (Center for Innovation Competence program, grant 03Z1HN31), the Vienna Center for Quantum Science and Technology (VCQ), the Austrian Nano-initiative Nanostructures of Atomic Physics (NAP-PLATON), and the Austrian Science Fund (FWF)
with the projects PhoQuSi Photonic Quantum Simulators (Y585-N20) and the doctoral programme CoQuS Complex Quantum Systems, the Vienna Science and Technology Fund (WWTF) under grant ICT12-041, and the Air Force Office of Scientific Research, Air Force Material Command, United States Air Force, under grant number FA8655-11-1-3004. B.C.S. acknowledges support from AITF (Alberta Innovates Technology Futures), NSERC (Natural Sciences and Engineering Research Council), and CIFAR (Canadian Institute for Advanced Research). The work of H.dG. is supported in part by NSERC of Canada. S.--H.T.: This material is based on research supported in part by the Singapore National Research Foundation under NRF Award No. NRF-NRFF2013-01. R.H., S.N. and A.S. acknowledge support from the German Ministry of Education and Research (Center for Innovation Competence programme, grant 03Z1HN31), the Deutsche Forschungsgemeinschaft (grant NO462/6-1), the Thuringian Ministry for Education, Science and Culture (Research group Spacetime, grant 11027-514).\\
The authors declare that they have no competing financial interests.\\
\end{acknowledgments}


\clearpage

\appendix

	
\begin{widetext}
\section*{Appendix}
\subsection{Two--photon non--classical interference}
Two photons injected into different inputs of an arbitrary beam splitter or a network built from arbitrary beam splitters and phase shifters will interfere non--classically~\cite{Hong1987,Fearn1989}. This input state can be expressed as,
\begin{equation}
\label{eq:2in}
\ket{11} = (\hat{A}^{\dagger}_1(\alpha_1)e^{i\omega_1\tau_1})(\hat{A}^{\dagger}_2(\alpha_2)e^{i\omega_2\tau_2})\ket{0},
\end{equation}
with 
\begin{equation}
\label{eq:Ai}
\hat{A}^{\dagger}_i(\alpha_i)=\int\limits_{0}^\infty d\omega_i \alpha_i(\omega_i)\hat{a}^{\dagger}_i(\omega_i) \ ,
\end{equation}
for~$A^{\dagger}_i(\alpha_i)$ a creation operator for a photon with spectral function
\begin{equation}
\label{eq:alpha}
	\left |\alpha(\omega_i)\right|^2= \frac{1}{\sqrt{2\pi}\sigma_i}\exp\left(-\frac{(\omega_i-\omega_{c,i})^2}{2\sigma_i^2}\right)
\end{equation}
centered at time~$\tau_i$. The frequency-mode creation operators on the right-hand side (RHS) of equation~(\ref{eq:Ai}) satisfy the commutator relation 
\begin{equation}
\label{comm}
	[\hat{a}_i(\omega),\hat{a}^{\dagger}_j(\omega')]=\delta_{ij}\delta(\omega-\omega')\mathds{1}
\end{equation}
with~$\mathds{1}$ the identity operator. This commutator relation also defines the photons' symmetry under permutation operations. For two photons it is sufficient to define their relative temporal delay as $\Delta\tau=\tau_1-\tau_2$. Only in the case of ideal bosonic particles exhibiting no modal mismatch and perfect temporal overlap, i.e. $\Delta\tau=0$, does the RHS of equation~(\ref{comm}) become the well--known bosonic commutator relation describing perfect symmetry under exchange. When the two--photon input state (see equation~(\ref{eq:2in})) is mixed via a transformation matrix $B=U_{2\times2}$ and projected on an output where the two photons exit in different modes, the output probability becomes,

\begin{align}
P_c(\Delta\tau)&=\int d \omega_1 \int d \omega_2 \left|\langle 11| \hat{B}^\dag\hat{a}^{\dagger}_1(\omega_1)\hat{a}^{\dagger}_2(\omega_2) |0 \rangle   \right|^2\\
		&=
\begin{pmatrix}
	{\rm per}(B) \\
	{\rm det}(B)
	\end{pmatrix}^{\mathlarger{\dagger}}
	\left[\frac{1}{2}
\begin{pmatrix}
	1 &\quad\,\, 0 \\
	0 &\quad\,\, 1
	\end{pmatrix}
+\frac{1}{2}\zeta\text{e}^{- \xi\Delta\tau^2}
\begin{pmatrix}
	1 &\quad\, 0 \\
	0 &\,\,\, -1
	\end{pmatrix}
	\right]
\begin{pmatrix}
	{\rm per}(B) \\
	{\rm det}(B)
	\end{pmatrix}\label{eq6}\\
	&= \boldsymbol{v_2}^{\dagger} \big[\hat{R}^{(2)}(\Delta\tau)\big]\boldsymbol{v_2}\label{eq7},
\end{align}

with
\begin{equation}
	\zeta=\frac{2\sigma_1\sigma_2}{\sigma_1^2+\sigma_2^2}\exp\left({-\frac{(\omega_{c,1}-\omega_{c,2})^2}{2(\sigma_1^2+\sigma_2^2)}}\right),\;
\xi=\frac{\sigma_1^2\sigma_2^2}{\sigma_1^2+\sigma_2^2}
\end{equation}
denoting factors arising from the spectral overlap integral and

\begin{equation}
	\boldsymbol{v_2}=\frac{1}{\sqrt{2}}\begin{pmatrix}
	{\rm per}(B) \\
	{\rm det}(B)
	\end{pmatrix}
\end{equation}
the new basis vector constituted by matrix functions of the scattering submatrix $T$. As a second--order correlation effect this non--classical interference is dependent on the permutational symmetry of the interfering wavefunctions also reflected in the basis vector $\boldsymbol{v_2}$. For the case of indistinguishable photons ($\omega_{c,1}=\omega_{c,2}$, $\sigma_1=\sigma_2$ or $\Delta\tau=0$), the output probability is only proportional to the permanent. This is a function symmetric under permutation of rows of the transformation matrix arising in photon interferometry due to bosonic exchange symmetry. However, with loss of complete indistinguishability ($\omega_{c1}\neq\omega_{c2}$, $\sigma_1\neq\sigma_2$ and $\Delta\tau\neq0$), equation~(\ref{eq6}) becomes proportional to a combination of the determinant and the permanent. This is a consequence of the input state losing its symmetry under exchange. Equation~(\ref{eq7}) decouples the influence of the interferometer from the influence of the input state. The latter is contained in the diagonal $2\times2$ rate--matrix $\hat{R}^{(2)}(\Delta\tau)$, whereas the description of the interferometer is absorbed in the new basis vector $\boldsymbol{v_2}$. The two non--zero entries of the rate--matrix, $\hat{R}^{(2)}_{11}$ and $\hat{R}^{(2)}_{22}$, are sufficient to reveal the nature of the non--classical interference of two photons of arbitrary coherence. Where $\hat{R}^{(2)}_{11}$ quantifies the contribution from the permanent of the scattering submatrix $\hat{R}^{(2)}_{22}$ quantifies the contribution from the determinant of the scattering submatrix. The output probability $P_c$ is recovered by calculting those matrix functions.\\

\subsection{Three--photon non--classical interference}
Non--classical interference of photons depends on indistinguishability of the interfering photons and transformations mixing the modes. Adding a third photon noticeably increases the complexity. An input state corresponding to three photons in three different transverse spatio--temporal modes can be described as

\begin{align}
\ket{111}=(A^{\dagger}_1(\alpha_1)e^{i\omega\tau_1})(A^{\dagger}_2(\alpha_2)e^{i\omega'\tau_2})(A^{\dagger}_3(\alpha_3)e^{i\omega''\tau_3})\ket{0}\ .
\end{align}

For three photons it is sufficient to define two relative temporal delays, $\Delta\tau_1=\tau_1-\tau_2$ and $\Delta\tau_2=\tau_3-\tau_2$. When this input state is transformed via a submatrix $T=U_{3\times3}$ and projected on an output where the three photons exit in different modes the fully expanded output probability can be written as

\begin{align}
P_{111}(\Delta\tau_1,\Delta\tau_2)=&\int d\omega\int d\omega' \int d
\omega''|\bra{111}\hat{T}^\dag a_1^\dag(\omega)a_2^\dag(\omega')a_3^
\dag(\omega'')\ket{0}|^2 \\
=&\label{fullterms}\frac{1}{6}|\Det(T)|^2 +\frac{2}{9}|\imm(T_{132})|^2+\frac{1}{9}\imm^*(T_{132})\imm(T_{213})+\frac{1}{9}\imm(T_{132})\imm^*(T_{213})\\ \nonumber
&+\frac{2}{9}|\imm(T_{213})|^2+\frac{2}{9}|\imm(T_{231})|^2+\frac{2}{9}|\imm(T)|^2+\frac{1}{9}\imm(T_{231})\imm^*(T)\\
&+\frac{1}{6}|\Per(T)|^2+\frac{1}{9}\imm(T)\imm^*(T_{231})\nonumber\\
& +\zeta_{13}\exp(-2\xi_{13}(\Delta\tau_1-\Delta\tau_2)^2)\Big( -\frac{1}{6}|\Det(T)|^2-\frac{2}{9}\imm(T)\imm^*(T_{132})-\frac{1}{9}\imm(T)\imm^*(T_{213})\nonumber\\ 
&-\frac{1}{9}\imm^*(T_{132})\imm(T_{231})+\frac{1}{9}\imm^*(T_{213})\imm(T_{231})-\frac{1}{9}\imm(T_{132})\imm^*(T_{231})\nonumber\\
&+\frac{1}{9}\imm(T_{213})\imm^*(T_{231})-\frac{2}{9}\imm(T_{132})\imm^*(R)-\frac{1}{9}\imm(T_{213})\imm^*(T)+\frac{1}{6}|\Per(T)|^2\Big )\nonumber\\
&+\zeta_{12}\exp(-2\xi_{12}\Delta\tau_1^2)\Big(-\frac{1}{6}|\Det(T)|^2+\frac{1}{9}\imm(T)\imm^*(T_{132})+\frac{2}{9}\imm(T)\imm^*(T_{213})\nonumber\\
&+\frac{2}{9}\imm^*(T_{132})\imm(T_{231})+\frac{1}{9}\imm^*(T_{213})\imm(T_{231})+\frac{2}{9}\imm(T_{132})\imm^*(T_{231})\nonumber\\
&+\frac{1}{9}\imm(T_{213})\imm^*(T_{231})+\frac{1}{9}\imm(T_{132})\imm^*(T)+\frac{2}{9}\imm(T_{213})\imm^*(T)+\frac{1}{6}|\Per(T)|^2\Big)\nonumber\\
&+\zeta_{23}\exp(-2\xi_{23}\Delta\tau_2^2)\Big(-\frac{1}{6}|\Det(T)|^2+\frac{1}{9}\imm(T)\imm^*(T_{132})-\frac{1}{9}\imm(T)\imm^*(T_{213})\label{tau4}\nonumber\\
&-\frac{1}{9}\imm^*(T_{132})\imm(T_{231})-\frac{2}{9}\imm^*(T_{213})\imm(T_{231})-\frac{1}{9}\imm(T_{132})\imm^*(T_{231})\nonumber\\
&-\frac{2}{9}\imm(T_{213})\imm^*(T_{231})+\frac{1}{9}\imm(T_{132})\imm^*(T)-\frac{1}{9}\imm(T_{213})\imm^*(T)+\frac{1}{6}|\Per(T)|^2\Big)\nonumber\\
&+\zeta_{123}\exp(-I_a+iI_s)\Big(\frac{1}{6}|\Det(T)|^2-\frac{1}{9}|\imm(T_{132})|^2-\frac{2}{9}\imm^*(T_{132})\imm(T_{213})\nonumber\\
&+\frac{1}{9}\imm(T_{132})\imm^*(T_{213})-\frac{1}{9}|\imm(T_{213})|^2+\frac{1}{9}\imm(T)\imm^*(T_{231})-\frac{1}{9}|\imm(T_{231})|^2\nonumber\\
&-\frac{1}{9}|\imm(T)|^2-\frac{2}{9}\imm(T_{231})\imm^*(T)+\frac{1}{6}|\Per(T)|^2\Big)\nonumber\\
&+\zeta_{123}\exp(-I_a-iI_s)\Big(\frac{1}{6}|\Det(T)|^2-\frac{1}{9}|\imm(T_{132})|^2-\frac{2}{9}\imm(T_{132})\imm^*(T_{213})\nonumber\\
&+\frac{1}{9}\imm^*(T_{132})\imm(T_{213})-\frac{1}{9}|\imm(T_{213})|^2+\frac{1}{9}\imm^*(T)\imm(T_{231})-\frac{1}{9}|\imm(T_{231})|^2\nonumber\\
&-\frac{1}{9}|\imm(T)|^2\nonumber-\frac{2}{9}\imm^*(T_{231})\imm(T)+\frac{1}{6}|\Per(T)|^2\Big)\ ,\nonumber
\end{align}
with
\begin{align}
	\zeta_{123}=&\sqrt{\zeta_{12}\zeta_{23}\zeta_{13}},
					\nonumber\\
	I_a\equiv & I_a(\Delta\tau_1,\Delta\tau_2)=-(\Delta \tau_1)^2 \frac{\xi_{12}}{2}-(\Delta \tau_1 - \Delta\tau_2)^2 \frac{\xi_{13}}{2}-(\Delta\tau_2)^2 \frac{\xi_{23}}{2},
					\nonumber\\
	I_s\equiv & I_s(\Delta\tau_1,\Delta\tau_2)=\Delta \tau_1 \nu_{12}-(\Delta\tau_1-\Delta\tau_2)\nu_{13}-\Delta\tau_2\nu_{23},
					\nonumber\\
		\zeta_{ij}=&\frac{2\sigma_i\sigma_j}{\sigma_i^2+\sigma_j^2}\exp\left(-\frac{(\omega_{c,i}-\omega_{c,j})^2}{2(\sigma_i^2+\sigma_j^2)}\right),\\
\xi_{ij}=&\frac{2\sigma_i^2\sigma_j^2}{\sigma_i^2+\sigma_j^2},\;
		\nu_{ij}=\frac{\omega_{c,i}\sigma_j^2+\omega_{c,j}\sigma_i^2}{\sigma_i^2+\sigma_j^2}.
\end{align}
The subscripts denote the mode labels for the submatrix~$T$. $T_{ijk}$ is the matrix $T$ with the rows permuted according to $1\rightarrow i$, $2\rightarrow j$, and $3\rightarrow k$.\\
For a more elegant expression, equation~(\ref{fullterms}) can be simplified introducting six matrices, $\openone$, $\rho_{12}$, $\rho_{13}$, $\rho_{23}$, $\rho_{123}$, and $\rho_{132}$:
\begin{align}
\label{eq:allsym}
	P_{111}(\Delta\tau_1,\Delta\tau_2)
		=&(\hat{P}\hat{S}\boldsymbol{v_3})^{\dagger}\big[\openone + \rho_{12}\zeta_{12}e^{-\xi_{12}\Delta\tau_1^2}+\rho_{23}\zeta_{23}e^{-\xi_{23}\Delta\tau_2^2}\nonumber\\
&+\rho_{13}\zeta_{13}e^{-\xi_{13}(\Delta\tau_1-\Delta\tau_2)^2}
+\zeta_{123}(\rho_{132}e^{\xi^*_{123}(\Delta\tau_1,\Delta\tau_2)}+\rho_{123}\text{e}^{\xi_{123}(\Delta\tau_1,\Delta\tau_2)})\big](\hat{P}\hat{S}\boldsymbol{v_3}) \\
= &\boldsymbol{v_3}^{\dagger}\big[\hat{R}^{(3)}(\Delta\tau_1,\Delta\tau_2)\big]\boldsymbol{v_3}\label{eq16},
\end{align}
where
\begin{equation}
	\xi_{123}(\Delta\tau_1,\Delta\tau_2)=I_a+iI_s.
\end{equation}
The vector $\hat{P}\hat{S}\boldsymbol{v_3}$ contains all the immanants and the determinant and permanent of $T$:
\begin{align}\label{eq:r}
	\hat{P}\hat{S}\boldsymbol{v_3}
		\equiv \begin{pmatrix}
	\frac{1}{\sqrt{6}}{\rm per}(T) \\
	\frac{1}{\sqrt{6}}{\rm det}(T)\\
	\frac{1}{2\sqrt{3}}{\rm imm}(T)+\frac{1}{2\sqrt{3}}{\rm imm}(T_{213})\\
	\frac{1}{6}{\rm imm}(T)-\frac{1}{3}{\rm imm}(T_{132})-\frac{1}{6}{\rm imm}(T_{213})+\frac{1}{3}{\rm imm}(T_{312})\\
	\frac{1}{6}{\rm imm}(T)+\frac{1}{3}{\rm imm}(T_{132})+\frac{1}{6}{\rm imm}(T_{213})+\frac{1}{3}{\rm imm}(T_{312})\\
	-\frac{1}{2\sqrt{3}}{\rm imm}(T)+\frac{1}{2\sqrt{3}}{\rm imm}(T_{213})
	\end{pmatrix}
\end{align}

with 

\begin{align}\label{eq:r2}
\boldsymbol{v_3}=\begin{pmatrix}
	{\rm per}(T) \\
	{\rm imm}(T) \\
	{\rm imm}(T_{132}) \\
	{\rm imm}(T_{213}) \\
	{\rm imm}(T_{312}) \\
	{\rm det}(T) 
\end{pmatrix},\ &
\hat{P}=\begin{pmatrix}
\frac{1}{\sqrt{6}} & \frac{1}{\sqrt{6}} & \frac{1}{\sqrt{6}} & \frac{1}{\sqrt{6}} & \frac{1}{\sqrt{6}} & \frac{1}{\sqrt{6}} \\
\frac{1}{\sqrt{6}} & -\frac{1}{\sqrt{6}} & -\frac{1}{\sqrt{6}} & \frac{1}{\sqrt{6}} & \frac{1}{\sqrt{6}} & -\frac{1}{\sqrt{6}} \\
\frac{1}{\sqrt{3}} & -\frac{1}{2\sqrt{3}} & \frac{1}{\sqrt{3}} & -\frac{1}{2\sqrt{3}} & -\frac{1}{2\sqrt{3}} & -\frac{1}{2\sqrt{3}} \\
0 & -\frac{1}{2} & 0 & -\frac{1}{2} & \frac{1}{2} & \frac{1}{2} \\
0 & \frac{1}{2} & 0 & -\frac{1}{2} & \frac{1}{2} & -\frac{1}{2} \\
-\frac{1}{\sqrt{3}} & -\frac{1}{2\sqrt{3}} & \frac{1}{\sqrt{3}} & \frac{1}{2\sqrt{3}} & \frac{1}{2\sqrt{3}} & -\frac{1}{2\sqrt{3}}
\end{pmatrix}, \nonumber 
\hat{S}=\begin{pmatrix}
\frac{1}{6} & \frac{1}{3} & 0 & 0 & 0 & \frac{1}{6} \\
\frac{1}{6} & 0 & \frac{1}{3} & 0 & 0 & -\frac{1}{6}\\
\frac{1}{6} & 0 & 0 & \frac{1}{3} & 0 & -\frac{1}{6} \\
\frac{1}{6} & -\frac{1}{3} & 0 & 0 & -\frac{1}{3} & \frac{1}{6} \\
\frac{1}{6} & 0 & 0 & 0 & \frac{1}{3} & \frac{1}{6} \\
\frac{1}{6} & 0 & -\frac{1}{3} & -\frac{1}{3} & 0 & -\frac{1}{6}
\end{pmatrix}.
\end{align}

Here $\hat{P}$ is a basis--transformation and $\hat{S}$ is a matrix mapping matrix--elements to matrix functions. The six matrices $\rho$ are, in fact, permutation matrices reduced to block--diagonal form:
\begin{align}
\openone=\begin{pmatrix}
	1 & 0 & 0 & 0 & 0 & 0 \\
	0 & 1 & 0 & 0 & 0 & 0\\
	0 & 0 & 1 & 0 & 0 & 0 \\
	0 & 0 & 0 & 1 & 0 & 0 \\
	0 & 0 & 0 & 0 & 1 & 0 \\
	0 & 0 & 0 & 0 & 0 & 1
\end{pmatrix},\ &
\rho_{12}=\begin{pmatrix}
1 & 0 & 0 & 0 & 0 & 0 \\
0 & -1 & 0 & 0 & 0 & 0\\
0 & 0 & 1 & 0 & 0 & 0 \\
0 & 0 & 0 & -1 & 0 & 0 \\
0 & 0 & 0 & 0 & 1 & 0 \\
0 & 0 & 0 & 0 & 0 & -1
\end{pmatrix}, \nonumber \\
\rho_{23}=\begin{pmatrix}
1 & 0 & 0 & 0 & 0 & 0 \\
0 & -1 & 0 & 0 & 0 & 0\\
0 & 0 & -\frac{1}{2} & -\frac{\sqrt{3}}{2} & 0 & 0 \\
0 & 0 & -\frac{\sqrt{3}}{2} & \frac{1}{2} & 0 & 0 \\
0 & 0 & 0 & 0 & -\frac{1}{2} & -\frac{\sqrt{3}}{2} \\
0 & 0 & 0 & 0 & -\frac{\sqrt{3}}{2} & \frac{1}{2}
\end{pmatrix},\ & \rho_{13}=\begin{pmatrix}
1 & 0 & 0 & 0 & 0 & 0 \\
0 & -1 & 0 & 0 & 0 & 0\\
0 & 0 & -\frac{1}{2} & \frac{\sqrt{3}}{2} & 0 & 0 \\
0 & 0 & \frac{\sqrt{3}}{2} & \frac{1}{2} & 0 & 0 \\
0 & 0 & 0 & 0 & -\frac{1}{2} & \frac{\sqrt{3}}{2} \\
0 & 0 & 0 & 0 & \frac{\sqrt{3}}{2} & \frac{1}{2}
\end{pmatrix},\nonumber \\
\rho_{123}=\begin{pmatrix}
1 & 0 & 0 & 0 & 0 & 0 \\
0 & 1 & 0 & 0 & 0 & 0\\
0 & 0 & -\frac{1}{2} & -\frac{\sqrt{3}}{2} & 0 & 0 \\
0 & 0 & \frac{\sqrt{3}}{2} & -\frac{1}{2} & 0 & 0 \\
0 & 0 & 0 & 0 & -\frac{1}{2} & -\frac{\sqrt{3}}{2} \\
0 & 0 & 0 & 0 & \frac{\sqrt{3}}{2} & -\frac{1}{2}
\end{pmatrix}, \ & \rho_{132}=\begin{pmatrix}
1 & 0 & 0 & 0 & 0 & 0 \\
0 & 1 & 0 & 0 & 0 & 0\\
0 & 0 & -\frac{1}{2} & \frac{\sqrt{3}}{2} & 0 & 0 \\
0 & 0 & -\frac{\sqrt{3}}{2} & -\frac{1}{2} & 0 & 0 \\
0 & 0 & 0 & 0 & -\frac{1}{2} & \frac{\sqrt{3}}{2} \\
0 & 0 & 0 & 0 & -\frac{\sqrt{3}}{2} & -\frac{1}{2}
\end{pmatrix}.
\end{align}
Equation~(\ref{eq:allsym}) describes the same features as equation~(\ref{fullterms}) but highlights the permutational options for three photons. It is given in a maximally decoupled basis, which allows for a compact notation. The terms originating from the overlap integrals ($\zeta$ terms and $\xi$ terms) contain all the information on the physical properties of the interfering photons. The effect of the permutation symmetry of the photons is included in the permutation matrices $\rho$. Equation~(\ref{eq16}) features a even further compressed notation and allows for an elegant interpretation. Where the block--diagonal $6\times6$ rate--matrix $\hat{R}^{(3)}(\Delta\tau_1,\Delta\tau_2)$ contains all the information on the permutational symmetry and non--classical interference itself the basis--vector $\boldsymbol{v_3}$ contains the information on the interferometer. Two entries of this rate--matrix are sufficient for an interpretation. $F_{per}=\hat{R}_{11}^{(3)}(\Delta\tau_1,\Delta\tau_2)$ quantifies the fraction of the output probability distribution proportional to the permanent and $F_{det}=\hat{R}_{66}^{(3)}(\Delta\tau_1,\Delta\tau_2)$ to the determinant of the submatrix $T$. The contribution proportional to immanants can also be explicitly calculated. When only interested in their overall contribution this is given as $F_{imm}=1-F_{per}-F_{det}$. In the extremal case when all the photons are indistinguishable,
i.e.,
\begin{equation}
	\omega_{c,1}=\omega_{c,2}=\omega_{c,3}=\omega_c,\,\,
	\sigma_1=\sigma_2=\sigma_3=\sigma,\,\,
	\Delta\tau_1=\Delta\tau_2=0,
\end{equation}
we have $\zeta_{ij}=1$, $\xi_{ij}=\sigma^2$ and $\nu_{ij}=\omega$
so the output probability reduces from a superposition of $60$ terms to just $P_{111}\rightarrow |\Per(T)|^2$.

\subsection{Five--photon non--classical interference}
The simulated data for a BosonSampling instance of five photons of arbitrary distinguishability injected into an interferometric network of nine modes, shown in figure \textbf{5} of the main manuscript, is calculated as outlined in the accompanying Mathematica program. The Mathematica notebook "5 photon rate matrix.nb" contains modules that are necessary to compute the interferometer--independent rate matrix. First the regular representation of elements in $S_5$ is computed. These are $120\times120$ matrices which represent permutations of five objects. In the rate matrix, these representations form a basis, each weighted by an integral that is the corresponding overlap integral of five photons with arbitrary distinguishability caused by spectral and temporal mode mismatch. From these regular repesentations and the overlap integrals, the interferometer independent rate matrix, $Rm$ is obtained. The basis vector of this rate matrix is constituted by matrix functions of the $5\times5$ scattering submatrix $T_5$. The first and second entries of the basis vector are chosen to be the permanent and determinant of $T_5$ respectively. The remaining 118 entries of the basis vector need to cover all five partitions of immanants of $S_5$. Each partition is constituted by a number of elements equal to its dimension squared. Those elements are the immanant of the scattering submatrix of this partition and the immanants of non--redundant permutations of the scattering submatrix of this partition. In this decomposition different partitions of immanants do not mix. Therefore the fraction of an output probability proportional to a specific partition of an immanant can be calculated independently. For example the block in the rate matrix corresponding to the \{2,2,1\} partition is a $25\times25$ matrix, $Rm_{\{2,2,1\}}$ ranging from $Rm_{3,3}$ to $Rm_{27,27}$. Consequently, the related elements of the basis vector run from row $3$ to row $27$ and form a basis vector, $\boldsymbol{v_{\{2,2,1\}}}$ for this subspace. The output probability of this subspace can be calculated as $P_{\{2,2,1\}}=\boldsymbol{v_{\{2,2,1\}}}^{\dagger} Rm_{\{2,2,1\}}\boldsymbol{v_{\{2,2,1\}}}$. Individual partitions of immanants have a direct mapping to different physical scenarios of non--classical interference. $P_{\{2,2,1\}}$ quantifies the fraction of the output probability that arises due to a case of non--classical interference of two pairs of indistinguishable photons (the two pairs are distinguishable to one another) and the transmission of a completely distinguishable photon.

\subsection{Matrix Reconstruction}
The fabrication of integrated photonic networks using a femtosecond--laser--direct--writing technology works with high precision and high stability. Discrete unitary operators acting on modes can be realized solely from beam splitters and phase shifters~\cite{Reck1994}. These networks are arranged like cascaded Mach--Zehnder interferometers shown in Fig. \ref{FIG:interferometer}. Notably though, even advanced writing precision can introduce small deviations from the initially targeted values of individual elements. In our case this writing precision is limited to around \SI{50}{\nano\meter} over the whole length of the waveguide (in this experiment \SI{10}{\centi\meter}). In a cascaded interferometric arrangement small deviations of individual elements may add up to a noticeable deviation in the overall transformation. The splitting ratio of individual directional couplers is set by their mode separation and coupling length. Both characteristic variables are three orders of magnitude bigger than the positioning precision and therefore unaffected by it. Unfortunately small length fluctuations due to the positioning precision can introduce unintended phase shifts. In the worst case, i.e. a phase shifter spanning the whole length of a waveguide, the resultant phase shifts can even reach $\pi/8$. The layout used for the interferometric networks reported here (see Fig. \ref{FIG:interferometer}) circumvents this worst case. Even if the unintended phase shifts are decreased by a factor of $3$ at least; their influence needs to be evaluated and the actually implemented unitary needs to be reconstructed. The characterization procedure we use builds on the one introduced in~\cite{Laing2012,Tillmann2013}. Two--photon states from a down--conversion source are injected into different modes of the optical network to be characterized. This in situ method allows for a characterization with states having the same physical properties, e.g. frequency and spectral shape, as used later in the experiment. 

\begin{figure}[ht!]
\includegraphics[width=0.5\textwidth]{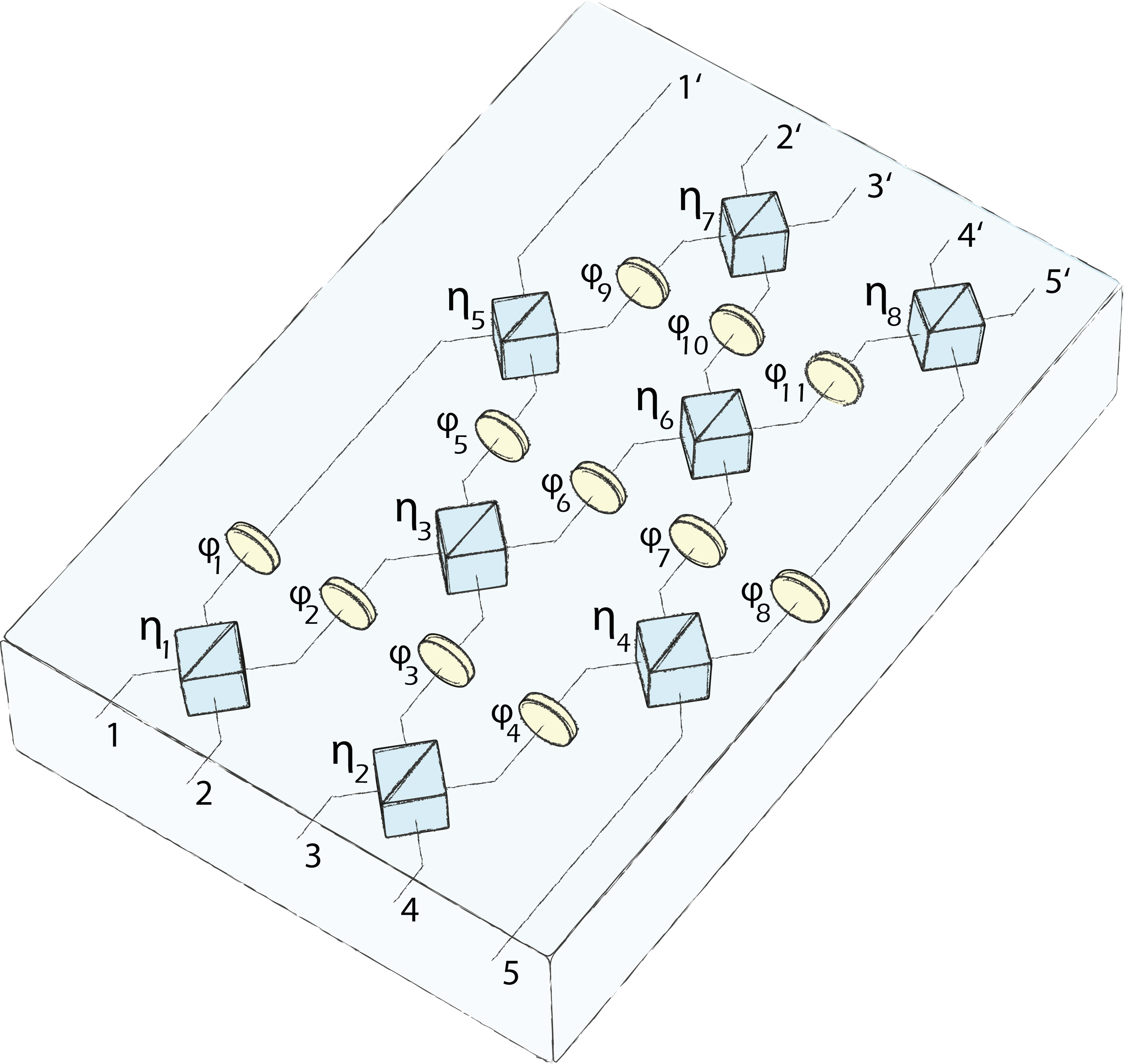}
\caption{\textbf{Integrated photonic network.} Schematic drawing of the optical network. The circuit consists of eight directional couplers ($\eta_1...\eta_8$), eleven phase shifters ($\phi_1...\phi_{11}$), five input modes (1...5) as well as of five output modes(1'...5'). To allow coupling to the waveguide with standard fiber--arrays the input and output modes are separated \SI{127}{\micro\meter} and the total length of the chip is \SI{10}{\centi\meter}.}
\label{FIG:interferometer}
\end{figure}

\subsubsection{Estimating the visibilities of submatrices}\label{Esti}
We assume the optical interferometer can be described by a $5\times 5$ unitary matrix and we reconstruct its transformation via visibilities measured by injecting two photons into any combination of two of its five inputs. The visibility for two photons entering input modes $i,j$ and exiting in the output modes $k,l$ can be calculated from the $2\times 2$ submatrix $U_{i,j,k,l}$. For five input and output modes this results in $\binom 5 2 \times \binom 5 2 = 100$ possibilities. Owing to the structure of the interferometer (see Fig.~\ref{FIG:interferometer}), a photon injected into port 5 cannot exit from output 1'. This leads to a visibility of zero for the four input pairs $ij=15,25,35,45$ and the output pairs $kl=15,25,35,45$. These visibilities are omitted from this reconstruction algorithm, so the unitary transformation is reconstructed from 84 non--zero visibilities.

Our interferometric network consists of eight beam splitters and eleven phase shifters. Each beam splitter implements a SU(2) transformation with matrix representation:
\begin{align}
\label{su2}
&\begin{pmatrix}
			\cos\frac{\beta}{2}
				& i\sin\frac{\beta}{2}\\
			i\sin\frac{\beta}{2}
				&\cos\frac{\beta}{2}
			\end{pmatrix} \ ,
\end{align}
where $\beta$ is the Euler angle associated with the transmittivity $\eta$ via the relationship $\eta=\cos^2(\beta/2)$. Note that in equation~(\ref{su2}) the beam splitter also implements a relative phase shift of $\pi$ between the first and second mode.\\ The eleven phase shifters produce additional phases in their respective modes. Each phase shifter has a matrix representation of
\begin{align}
\label{su2p}
&\begin{pmatrix}
			\text{e}^{i\alpha_1}
				& \text{0}\\
			\text{0}
				&\text{e}^{i\alpha_2}
			\end{pmatrix} \ ,
\end{align}
with $\alpha_i$ the phase shift in mode $i$.\\
The spectral shape of the photons is measured with a single--photon spectrometer (Ocean Optics QE6500) and to a good approximation is of Gaussian shape. Such Gaussians are defined by only two parameters, namely their central frequency and the variance, which for the $i^\text{th}$ photon of the input pair is given by equation~(\ref{eq:alpha}), and expressed here as
\begin{equation}
	\left|\phi_i(w)\right|^2=\frac{1}{\sqrt{2\pi}\sigma_i}\exp\left(-\frac{(\omega-\omega_{c,i})^2}{2\sigma_i^2}\right),\;
	i=1,\ 2.
\end{equation}
Assuming both photons exhibit identical spectral function, i.e. $|\phi_1(\omega)|^2=|\phi_2(\omega)|^2$, and the detectors are modeled by the detection positive--operator valued measure (POVM) with two elements $\{\Pi_0,\Pi_1\}$ satisfying completeness, $\sum_i \Pi_i=\mathbb{I}$,
\begin{equation}
\Pi_1=\int d\omega a^\dag(\omega)\ket{0}\bra{0}a(\omega) \ , \Pi_0=\mathbb{I}-\Pi_1 \ ,
\end{equation}
then the visibility is
\begin{align}\label{vis}
V=-\frac{h_1 h_2^*+h_1^* h_2 }{|h_1|^2+|h_2|^2} \ ,
\end{align}
with
\begin{equation}
	h_1=U^{11}_{i,j,k,l}U^{22}_{i,j,k,l},\;h_2=U^{12}_{i,j,k,l}U^{21}_{i,j,k,l} \ ,
\end{equation}
and $U^{a,b}_{i,j,k,l}$ denotes the element in the $a^{\rm th}$ row and $b^{\rm th}$ column of the matrix $U_{i,j,k,l}$. In an experiment the two photons will always have slightly different spectral functions whose mismatch needs to be accounted for. The central wavelengths and spectral bandwidths of the photons used in this characterization measurement are $\lambda_{c,1}=$\,\SI{789.05}{\nano\meter}, $\Delta \lambda_1=$\,\SI{2.9}{\nano\meter}, and $\lambda_{c,2}=$\,\SI{788.60}{\nano\meter}, $\Delta\lambda_{2}=$\,\SI{2.9}{\nano\meter} respectively. The coincidence counts $N_{c}$ as a function of time delay $t$ and spectral mode mismatch are

\begin{align}\label{coinccount}
N_c(t)=(1+T*t)(Y_0+A\frac{2\sigma_1\sigma_2}{\sigma_1^2+\sigma_2^2}\exp\left(-\frac{(\omega_{c,1}-\omega_{c,2})^2+4\sigma_1^2\sigma_2^2(t-t_c)^2}{2(\sigma_1^2+\sigma_2^2)}\right)-(HO_1+HO_2-d)) \ ,
\end{align}
where $Y_0$, $A$, $t_c$ and $T$ are parameters to be fitted to the experimental data. The experimental data for a given input/output combination $i,j,k,l$ it is typically recorded for 30 increments with a stepwidth of \SI{66}{\femto\second} and integrated over \SI{800}{\second} each step. The coincidences are read out by a field--programmable gate array logic (FPGA logic). As individual delays are set by translating a fiber coupler with a motorized screw (Newport LTA--HL) there can be a small drift in coupling efficiency over the whole delay--range of \SI{2000}{\femto\second}. Without this drift, the background of the visibility would be a horizontal straight line. For drifts smaller than $5\,\%$ of the two--photon flux the drift is in good approximation linear and can be modelled with an additional parameter, $T$. The positioning precision of the delay lines is limited to approximately $\pm$\,\SI{5}{\micro\meter} which is within $2.5\,\%$ of the coherence time of the interfering photons. When the two--photon input state is generated via down--conversion pumped by a pulsed laser system, higher order emission can lead to unwanted contribution to the input state. The first higher order, which is a four--fold emission, causes a small contribution of two photons in each input mode during the characterization of a $2\times2$ submatrix.
This can add a constant background to the two--fold coincidences in the following scenario: two photons in one input mode are lost and the two photons in the other input mode leave the network in different output ports. We measure such contributions by blocking one of the two input--modes and recording the two--photon coincidences at the output. These signals are labelled $HO_1$ and $HO_2$ respectively and subtracted from the data.
The background coincidence rate $d$ may be interpreted as a contribution to $N_c$ stemming from dark counts due to electrical noise and background light. This rate $d$ is also present in $HO_1$ and $HO_2$. Therefore it has to be added to equation~(\ref{coinccount}) to account for all unwanted coincidences only once. The error for the raw data was verified to be Poissonian. For the data processing the error of the higher order term $(HO_1+HO_2-d)$ and the abscissa--error caused by the limited alignment precision of the delay lines need to be taken into account additionally. These errors provide weighting in the minimization algorithm and influence the standard errors of the fitted parameters. The visibility, 
\begin{align}
V=1-\frac{Y_0+A}{Y_0} \ ,
\end{align}
is finally calculated from the parameters $Y_0$ and $A$, whereas the width of the dip or peak is fixed by the spectral function of the two photons. \textcolor{black}{Only 84 out of 100 visibilities are non--zero and their value, $\{V_i^{\rm(exp)}, i=1,\ldots,84 \}$ and standard deviation, $\{\sigma_i,i=1,\ldots, 84\}$ are extracted via the procedure outlined above. The resultant data fit with theory exhibits $\chi^2_\text{reduced}=1.74$\cite{taylor1997introduction}. An example for one of the 84 datasets is shown in Fig.~\ref{FIG:2hvexample}}.

\begin{figure}[ht!]
\includegraphics[width=0.8\textwidth]{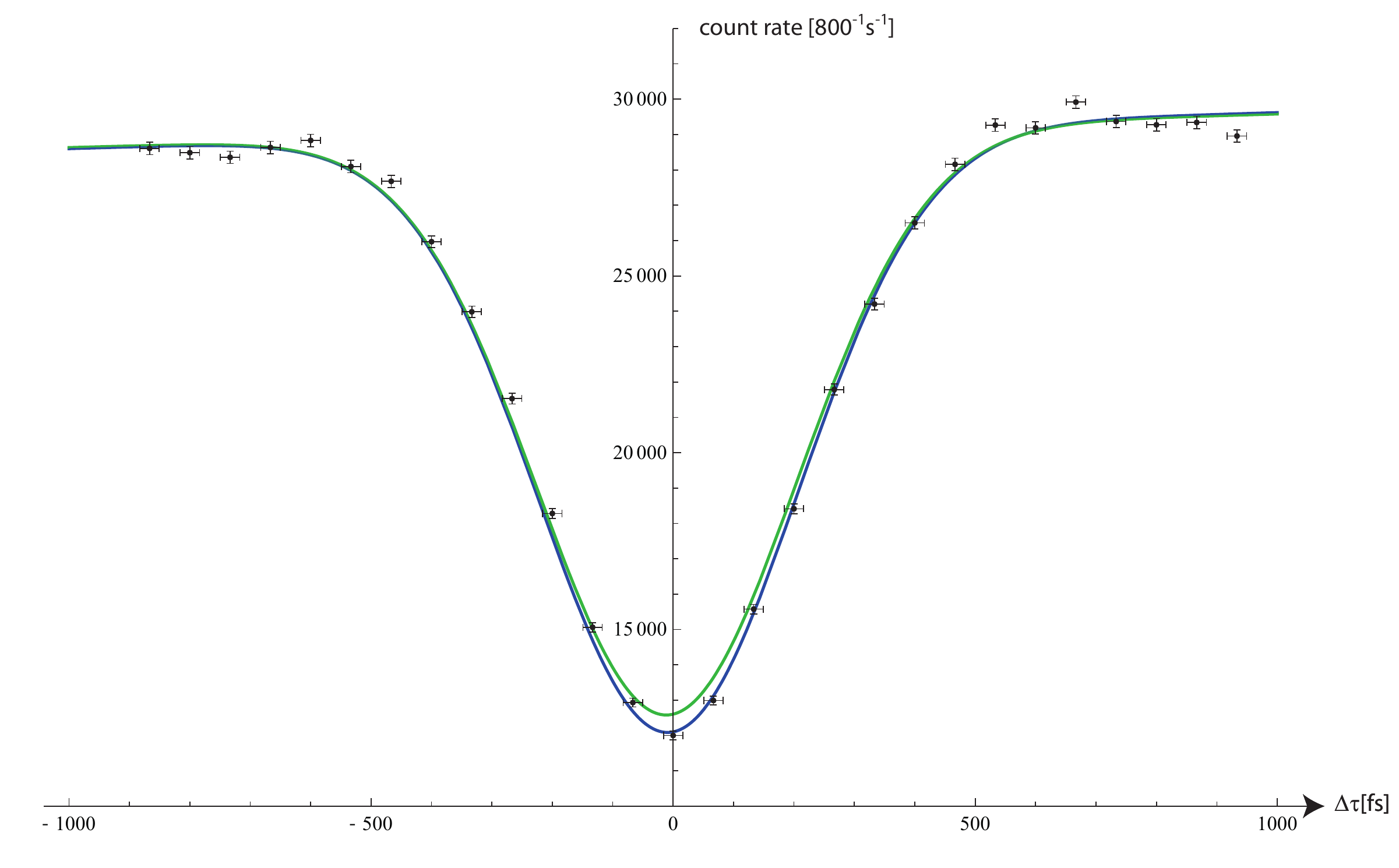}
\caption{\textbf{Example for one dataset used for the reconstruction of $U_5$.} \textcolor{black}{The best fit of equation~(\ref{coinccount}) to the data--set is shown in blue. Here the visibility is calculated from the best fit parameters $Y_0$ and $A$. The reduced $\chi^2$ resulting from the fit shown in blue is $\chi_{\rm blue}^2=2.02$. Fluctuations in the count rate for values of $|\Delta\tau|>$\SI{500}{\femto\second} drive the reduced $\chi^2$ away from $1$. These fluctuations can be interpreted as the random noise background in the lab and the increased reduced $\chi^2$ is reflecting that. However, the precision of the fitted parameters $Y_0$ and $A$ and ultimately the extracted visibility $V$ is only marginally affected by these fluctuations. The $5\times5$ unitary description of the interferometric network is reconstructed from 84 of these visibilities. The curve in green, $N_c^{\rm (th)}(t)$ (see equation~(\ref{Nc(th)})) is calculated from four matrix entries of the reconstructed unitary and results in an overlap with the data of $\chi_{\rm green}^2=2.70$.} \textcolor{black}{The agreement of these two curves and corresponding reduced $\chi^2$s is a qualitative measure for the precision of the reconstruction.}}
\label{FIG:2hvexample}
\end{figure}

\subsubsection{Parameter estimation and reconstruction of the unitary matrix}\label{ParaEsti}
\textcolor{black}{A unitary transformation of a linear optical network can be reconstructed from single--photon transmission probabilities and two--photon interference visibilities\cite{Laing2012}. A technique using coherent states\cite{Rahimi-Keshari2013} follows a similar approach. Both techniques reconstruct the unitary description in a dephased representation where the single--photon or single--input coherent state data is used to estimate the real parts of the matrix--entries. The imaginary parts of the matrix--entries are reconstructed from the two--photon interference visibilities or directly from the relative phase shifts. When the layout and initially targeted parameters of the building blocks (see Fig.~\ref{FIG:interferometer}) of the interferometric network are known, their actual parameters can be fitted alternatively. Our technique uses an over--complete set of visibilities and the} parameters of the interferometer that give an optimal fit to the experimentally measured visibilities are obtained using a \textcolor{black}{least square optimization weighted with the standard errors of the experimental visibilities (see~\ref{Esti} for details).} Eight of the $19$ parameters are transmittivities, $\beta_1,\beta_2,\ldots \beta_8$, and eleven are phases, $\phi_1,\phi_2,\ldots \phi_{11}$. To find the best--fit set of parameters, the data was processed with a Matlab program that uses fmincon to minimize the function $V_{\rm opt}$,
\begin{align}\label{approxchisq}
V_{\rm opt}=\sum_{i=1}^{84}\frac{\left(V_i^{\rm (exp)}-V_i^{\rm (th)}\right)^2}{\sigma_i^2\Gamma}\ ,
\end{align}
where $V_i^{\rm (th)}$ is the theoretical value of the visibility calculated from our special unitary model of the interferometer using equation~(\ref{vis}) for the $i^{\rm th}$ data set, and $\Gamma$ is a constant value equal to $({\rm number\,of\, data\,sets\,in\,visibilities} - {\rm number\,of\,parameters} - 1)=2522-19-188-1=2314$. \textcolor{black}{Equation~(\ref{approxchisq}) looks similar to a reduced $\chi^2$ but has to be interpreted differently. A value close to $0$ is desirable and indicates good agreement between experimentally extracted and theoretically predicted visibilities. In our case the result is $V_\text{opt}=0.351$.}\\

The $5\times 5$ reconstructed matrix $U_5$ using the procedure outlined above is 

\begin{align}
	U_5
		=\begin{pmatrix}
			0.0320-0.3370 i & 0.07239+0.8203 i & -0.2780-0.1060 i & 0.1228-0.3220 i & 0\\
			0.0114+0.2751 i & -0.3863+0.1860 i & -0.1353+0.2073 i & -0.7842-0.1502 i & 0.0124 - 0.2036 i \\
			-0.7757-0.2328 i & -0.2937+0.0018i & -0.2677-0.0162i & 0.0267+0.3517i & -0.2476-0.0151 i\\
			0.1444-0.2611 i & -0.1518-0.0840 i & -0.1392+0.0839i & -0.1327-0.0092i & 0.0203+0.8449i\\
			0.2225+0.1231i & 0.0715-0.1293i & -0.7929-0.0268i & 0.0871+0.3067i & 0.4123-0.1121i
		\end{pmatrix}.
\end{align}
\subsection{Quality of the reconstructed description}
Using this matrix, the probability of coincidence counts, $P_{11}^{\rm (th)}$, can be predicted for any two--photon inputs and outputs. For the inputs $i$ and $j$, $i<j$ and outputs $k$ and $l$, $k<l$, this reads as
\begin{align}
P_{11}^{\rm (th)}(t-t_c)=|U^{ki}_{5}U^{jl}_{5}|^2+|U^{li}_{5}U^{kj}_{5}|^2+(U^{li}_{5}U^{kj}_{5}{U^{ki}_{5}}^*{U^{jl}_{5}}^*+{U^{li}_{5}}^* {U^{kj}_{5}}^* U^{ki}_{5}U^{jl}_{5})f(t-t_c) \ ,
\end{align}
where
\begin{align}
f(t)\equiv(2\sigma_1\sigma_2/(\sigma_1^2+\sigma_2^2))\exp\left(-\frac{(\omega_{c,1}-\omega_{c,2})^2+4\sigma_1^2\sigma_2^2t^2}{2(\sigma_1^2+\sigma_2^2)}\right),
\end{align}
and $U_5^{ab}$ is the element in the $a^{\rm th}$ row and $b^{\rm th}$ column of $U_5$. The actual coincidence count is then
\begin{align}\label{Nc(th)}
N_c^{\rm (th)}(t)=N_0 (1+T) P_{11}^{\rm (th)}(t-t_c),
\end{align}
where $N_0$, $t_c$ and $T$ are parameters used to find the best fit to the experimental data. The exact $\chi^2_{\rm red}$ is calculated using
\begin{align}
\chi^2_{\rm red}=\sum_{i=1}^m \frac{\left(N_{c,i}^{\rm(exp)}-N_{c,i}^{\rm (th)}\right)^2}{\nu\epsilon_i^2} \ ,
\end{align}
where $m=3030$, $\nu=m-20-100-1=2909$, $\epsilon_i$ is the error for the corresponding datapoint, and $N_{c,i}^{\rm(exp)}$ denoting the experimental data corrected for higher order emissions. The sum is taken over the data set and the index labels the data. The obtained $\chi^2_{\rm red}$ between the data and the predicted coincidence counts using $U_5$ is
\begin{align}\label{result}
\chi^2_{\rm red}=2.086 \ .
\end{align}
\textcolor{black}{This value should be compared to reduced $\chi_{\rm exp}^2=1.74$ obtained by fitting the primary data to extract the $84$ visibilities in the beginning (see \ref{Esti}). The difference between those two reduced $\chi^2$s can be attributed to accuracy of the reconstructed unitary matrix $U_5$. While fluctuations in the count rate for values of $|\Delta\tau|>$\SI{500}{\femto\second} drive both reduced $\chi^2$s away from $1$, the difference between the two reduced $\chi^2$s is relatively small ($\approx 0.35$). An example for one of the 100 data sets is shown in Fig.~\ref{FIG:2hvexample}.}
 
\subsection{State Generation}
We use an \SI{80}{\mega\hertz} Ti:Sapphire oscillator emitting \SI{150}{\femto\second} pulses at a wavelength of \SI{789}{\nano\meter} which get frequency doubled via a $LiB_3O_5$ (LBO). The upconverted beam is focused into a \SI{2}{\milli\meter} thick $\beta-BaB_2O_2$ (BBO) crystal cut for degenerate non--collinear type--II spontaneous parametric down--conversion. To achieve near spectral indistinguishability and enhance temporal coherence of the down--converted wave packets the photons are filterd by $\lambda_{\mathrm{FWHM}} = $\SI{3}{\nano\meter} interference filters.
The source is aligned to emit the maximally entangled state
\begin{equation}
\ket{\phi^+}=\frac{1}{\sqrt{2}}\left(\ket{H}_a\ket{H}_b+\ket{V}_a\ket{V}_b\right),
\end{equation}
when pumped with low pump power (\SI{200}{\milli\watt} cw--equivalent). H and V denote horizontal and vertical polarization and a and b are the two spatial emission--modes. When pumped with higher pump powers (\SI{700}{\milli\watt} cw--equivalent) noticeable higher order emission occurs:
\begin{equation}
\ket{\psi}_{a,b}=\frac{1}{\sqrt{3}}(\ket{HH}_a\ket{HH}_b+\ket{HV}_a\ket{HV}_b+\ket{VV}_a\ket{VV}_b).
\end{equation}
This state is guided to two polarizing beam splitter (PBS) cubes. A detection event in the trigger mode $a''$ heralds the generation of either the state $\ket{V}_{a'}\ket{V}_{b'}\ket{H}_{b''}$ or $\ket{HH}_{b''}$ (see Fig.~\ref{FIG:Chip}). Only in the first case are the three modes $a'$, $b'$, and $b''$ occupied with one single photon, whereas in the latter case mode $b''$ is occupied with two photons and mode $b'$ with vacuum. Post--selection on a four--fold coincidence between mode $a''$, $a'$, $b'$, and $b''$ allows for the heralding of the desired input state where only one photons enters each input mode. The half--wave plates in mode $a'$ and $b'$ are set to $45^{\circ}$ to render them indistinguishable in polarization from the other photons. This heralding scheme holds independently of any transformation for the photons in mode $a'$, $b'$, and $b''$ as long as it acts on spatial modes, e.g. consisting of beam splitters and phase shifters only.

\begin{figure}[ht!]
\includegraphics[width=0.7\textwidth]{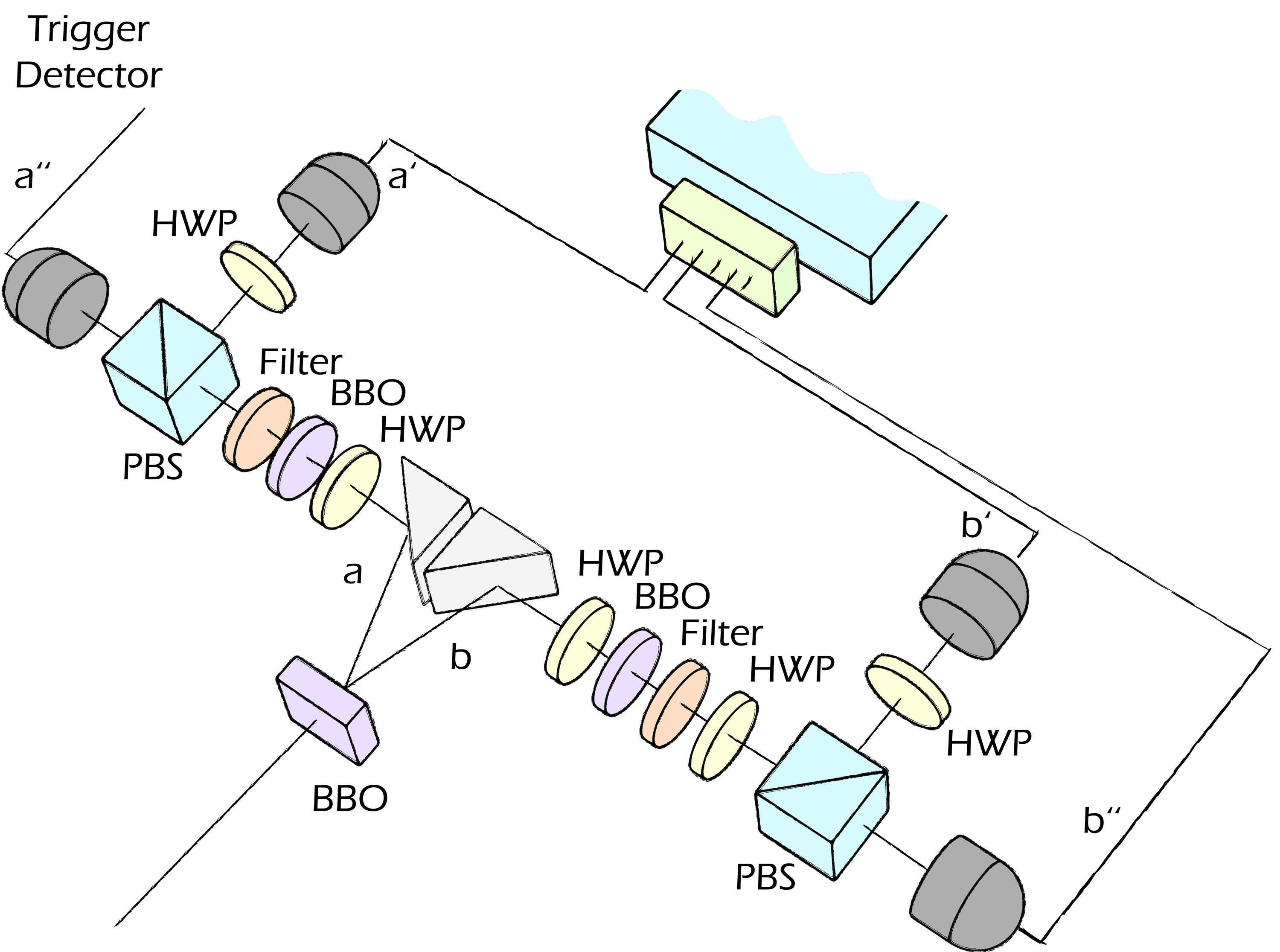}
\caption{\textbf{State generation.} A pump beam is focused into a \SI{2}{\milli\meter} $\beta$--$BaB_2O_2$ (BBO) crystal cut for non--collinear, degenerate, type--II down--conversion. The generated state is emitted into the spatial modes $a$ and $b$. A compensation scheme consisting of half--wave plates (HWPs) and \SI{1}{\milli\meter} thick BBO crystals is applied for countering temporal and spatial walk--off. Narrowband interference filters ($\lambda_{\text{FWHM}}=$\,\SI{3}{\nano\meter}) are applied to increase the temporal coherence of the photons and render them close to spectral indistinguishability. The modes $a$ and $b$ are subsequently split by polarizing beam splitter cubes (PBS) and two half--wave plates in their reflected ports are set to $45^{\circ}$ to ensure the same polarization in all four output modes ($a''$, $a'$, $b'$, and $b''$). With this scheme three indistinguishable photons in mode $a'$, $b'$, and $b''$ each can be heralded from a four--fold emission by a successful trigger event in mode $a''$.}
\label{FIG:Chip}
\end{figure}

\subsection{Analysis of the three--fold coincidence data}

Three photons are inserted into input modes 1,2 and 4 of the interferometric network. The spectral characteristics of these photons were measured using a single--photon spectrometer (Ocean Optics QE6500) and are in good approximation of Gaussian shape. Note that this spectral data differs slightly compared to the characterization measurements (see~\ref{Esti}).   

\begin{center}
\begin{tabular}{ |l|l|l| }
\hline
 & $\lambda_c$ & $\Delta \lambda_{\rm {FWHM}}$ \\ \hline
In1 & 789.35 nm & 2.85 nm \\ \hline
In2 & 789.52 nm & 2.79 nm \\ \hline
In4 & 789.41 nm & 2.72 nm \\ 
\hline
\end{tabular}
\end{center}

This spectral data allows to express the mode overlap integrals in dependence of the time delays $\Delta\tau_1$ and $\Delta\tau_2$ between the first and second photon and the second and third photon respectively. The theoretical prediction for the output probability in any of the ten three--fold output ports is then calculated using equation~(\ref{eq:allsym}). Consequently each $3\times3$ submatrix $R$ is constituted by matrix elements selected by the input and output ports. The output probability (see equation~(\ref{eq:allsym})) of any landscape contains a constant term and four terms proportional to different mode overlap functions. By sampling six points of pairwise temporal delay of $\Delta\tau_1$ and $\Delta\tau_2$ contributions of each of these terms can be assessed. These six points are 

\begin{center}
\begin{tabular}{ |l|l|l| }
\hline
 & $\Delta\tau_1$ & $\Delta\tau_2$ \\ \hline
P1 & 0\,fs & 130\,fs \\ \hline
P2 & 0\,fs & -870\,fs \\ \hline
P3 & -300\,fs & -170\,fs \\ \hline
P4 & -1000\,fs & -870\,fs \\ \hline
P5 & -1000\,fs & 130\,fs \\ \hline
P6 & -1000\,fs & 1130\,fs \\ 
\hline
\end{tabular}
\end{center}

An offset of $\Delta\tau_{\rm off}=\text{\SI{130}{\femto\second}}$ is introduced in the temporal delay mode $\Delta\tau_2$, otherwise the delays are set to combinations of \SI{0}{\femto\second}, \SI{-300}{\femto\second} and $\pm$\SI{1000}{\femto\second}. Precision of the temporal alignment was estimated to be $\pm$\SI{16}{\femto\second}. In one measurement run the points P1 to P6 are recorded consecutively for two hours each. To account for effects of drift this order is reversed in the next measurement run, therefore the points are recorded in the order P6 to P1. The four--fold count rates range from \SI{1}{\milli\hertz} to \SI{100}{\milli\hertz} dependent on the output combination. In between each measurement run the setup was realigned to optimize for maximal count rates. In order to obtain sufficient statistics, the whole data acquisition is repeated over 19 measurement runs for a total of 228 hours. \\
As Poissonian error modeling results in too optimistic error bars in case of long data acquisition due to multiple sources of error, we adapted the error modeling. The 19 measurements are independent runs therefore mean and standard deviation of the mean provide more useful information. Each individual measurement run is represented as a six--dimensional vector, with the $i^{\rm th}$ entry of the vector containing the four--fold counts of the $P_{i^{\rm th}}$ delay point integrated over two hours. These vectors can then be normalized to unit vectors thereby obtaining relative output probabilities. The mean and the standard deviation of the mean can now be calculated for each of the six delay points. Ultimately the overlap with the theoretical prediction is obtained by a least squared minimization weighted with the standard deviations. Here a linear scaling factor is introduced relating the relative experimental probabilities to the absolute theoretical ones. The goodness of fit is calculated using the reduced $\chi^2$. The number of degrees of freedom is in this case $\nu=6-2=4$.
\\

The experimental data for the four different scenarios of BosonSampling affected by distinguishability, shown in figure \textbf{4} of the main manuscript, are recorded using the same method as above. The experimental data and theoretical prediction is contained in table~\ref{figure4data}.

\begin{table}
\begin{tabular}{ |l|l|l|l|l|l|l| }
\hline
\multicolumn{7}{ |c| }{Data for In124} \\
\hline
  & $\tau_1$ & $\tau_2$ & theory & experimental & red. $\chi^2$ & count rate\\ \hline
\multirow{6}{*}{Out134} 
  & 0 fs & 130 fs &3.41\%&$3.17\% \pm $ 0.26\%& &\\
  & 0 fs & -870 fs &1.89\%&$2.18\% \pm $ 0.19\%&&\\
  & -300 fs & -170 fs &3.13\%&$2.99\% \pm $ 0.25\%&$1.38$ & $\approx$ 10 mHz\\
	& -1000 fs & -870 fs &2.95\%&$2.96\% \pm $ 0.26\%&&\\
	& -1000 fs & 130 fs &2.20\%&$2.51\% \pm $ 0.21\%&&\\
  & -1000 fs & 1130 fs &2.73\%&$2.72\% \pm $ 0.31\%&&\\ \hline
\multirow{6}{*}{Out345} 
  & 0 fs & 130 fs &14.19\%&$14.73\% \pm $ 0.93\%& &\\
  & 0 fs & -870 fs &23.69\%&$24.01\% \pm $ 0.84\%&&\\
  & -300 fs & -170 fs &17.67\%&$19.1\% \pm $ 0.98\%&$1.10$ & $\approx$ 80 mHz\\
	& -1000 fs & -870 fs &25.09\%&$24.01\% \pm $ 0.85\%&&\\
	& -1000 fs & 130 fs &21.14\%&$21.32\% \pm $ 0.80\%&&\\
  & -1000 fs & 1130 fs &31.40\%&$30.85\% \pm $ 1.44\%&&\\ \hline
\end{tabular}
\caption{data for the coincidence landscapes of figure \textbf{3}}
\label{3folds-data}
\end{table}

\begin{table}
\begin{tabular}{ |c|c|l|l|l|l|l| }
\hline
\multicolumn{7}{ |c| }{Data for figure 4} \\
\hline
  figure & $T_{ijk}$ & exp in \%& per in \% & imm in \%& det in \%& theo in \%\\
	\hline
\multirow{4}{*}{\textbf{4a}}
  & 245 & $1.46 \pm 0.39$ &1.72& 0.13 & 0.00 & 1.86\\
  & 235 & $10.02 \pm 0.83$ &11.32& 0.44 & 0.00 & 11.76\\
  & 123 & $46.75 \pm 2.95$ &33.38& 11.97& 0.00 & 45.36\\
	& 345 & $0.47 \pm 0.20$ &0.03& 0.16& 0.00 & 0.19\\
\multirow{2}{*}{$\Delta\tau_1=0$fs}
	& 234 & $7.24 \pm 0.80$ &7.08& 0.77 & 0.00 & 7.85\\
  & 134 & $6.69 \pm 0.71$ &6.30& 1.54 & 0.00 & 7.85\\
\multirow{2}{*}{$\Delta\tau_2=130$fs}
	& 125 & $7.96 \pm 0.89$ &5.21& 2.87 & 0.00 & 8.08\\
  & 145 & $1.69 \pm 0.40$ &1.41& 0.13 & 0.00 & 1.55\\	
	& 135 & $8.01 \pm 0.77$ &3.98& 1.10 & 0.00 & 5.08\\
  & 124 & $9.71 \pm 0.94$ &8.95& 1.50 & 0.00 & 10.45\\
	\hline
\multirow{4}{*}{\textbf{4b}}
  & 245 & $1.35 \pm 0.24$ &0.93& 0.59 & 0.01 & 1.53\\
  & 235 & $7.93 \pm 0.68$ &6.11& 2.45 & 0.11 & 8.67\\
  & 123 & $50.86 \pm 2.60$ &18.02& 29.62& 1.19 & 48.82\\
	& 345 & $0.78 \pm 0.16$ &0.02& 0.64& 0.01 & 0.66\\
\multirow{2}{*}{$\Delta\tau_1=-300$fs}
	& 234 & $6.00 \pm 0.41$ &3.82& 3.04 & 0.03 & 6.89\\
  & 134 & $5.12 \pm 0.58$ &3.40& 1.21 & 0.01 & 4.61\\
\multirow{2}{*}{$\Delta\tau_2=-170$fs}
	& 125 & $8.07 \pm 0.74$ &2.81& 3.69 & 0.03 & 6.53\\
  & 145 & $1.86 \pm 0.26$ &0.76& 1.20 & 0.02 & 1.98\\	
	& 135 & $8.64 \pm 0.68$ &2.15& 8.86 & 0.14 & 11.15\\
  & 124 & $9.39 \pm 0.59$ &4.83& 4.16 & 0.16 & 9.15\\
	\hline
	\multirow{4}{*}{\textbf{4c}}
  & 245 & $1.17\pm0.26$ &0.37& 0.77 & 0.05 & 1.19\\
  & 235 & $6.59\pm0.58$ &2.46& 3.35 & 0.59 & 6.40\\
  & 123 & $53.64\pm1.90$ &7.26& 40.72& 6.54 & 54.52\\
	& 345 & $0.92\pm0.20$ &0.01& 0.96& 0.03 & 1.00\\
\multirow{2}{*}{$\Delta\tau_1=-1000$fs}
	& 234 & $5.29\pm0.43$ &1.54& 3.63 & 0.15 & 5.32\\
  & 134 & $4.06\pm0.40$ &1.37& 2.43 & 0.04 & 3.84\\
\multirow{2}{*}{$\Delta\tau_2=-870$fs}
	& 125 & $8.19\pm0.64$ &1.13& 6.26 & 0.19 & 7.58\\
  & 145 & $1.57\pm0.22$ &0.31& 1.24 & 0.12& 1.67\\	
	& 135 & $9.91\pm0.74$ &0.87& 8.03 & 0.78 & 9.68\\
  & 124 & $8.67\pm1.02$ &1.95& 5.97 & 0.89 & 8.80\\
	\hline
	\multirow{4}{*}{\textbf{4d}}
  & 245 & $0.92\pm0.23$ &0.19& 0.51 & 0.17 & 0.86\\
  & 235 & $5.12\pm0.58$ &1.21& 1.91 & 1.97 & 5.09\\
  & 123 & $58.26\pm2.73$ &3.58& 33.36& 21.69 & 58.63\\
	& 345 & $0.60\pm0.09$ &0.00& 0.56& 0.11 & 0.68\\
\multirow{2}{*}{$\Delta\tau_1=-1000$fs}
	& 234 & $4.17\pm0.39$ &0.76& 2.85 & 0.50 & 4.11\\
  & 134 & $4.03\pm0.51$ &0.68& 2.79 & 0.12 & 3.59\\
\multirow{2}{*}{$\Delta\tau_2=1130$fs}
	& 125 & $7.38\pm0.62$ &0.56& 6.00 & 0.64 & 7.19\\
  & 145 & $1.32\pm0.28$ &0.15& 0.91 & 0.39 & 1.45\\	
	& 135 & $10.00\pm0.38$ &0.43& 7.15 & 2.57 & 10.15\\
  & 124 & $8.18\pm0.71$ &0.96& 4.33 & 2.95 & 8.25\\
	\hline

\end{tabular}
\caption{data for the normalized output probability distributions of figure \textbf{4}}
\label{figure4data}
\end{table}
\end{widetext}
\clearpage

\newpage


\end{document}